\documentclass[parskip=half,american]{scrartcl}
\usepackage[T1]{fontenc}
\usepackage{textcomp}
\usepackage[utf8]{inputenc}
\usepackage{xcolor}
\usepackage{babel}
\usepackage{array}
\usepackage{url}
\usepackage{multirow}
\usepackage{varwidth}
\usepackage{footnotehyper}
\usepackage{amsmath}
\usepackage{amssymb}
\usepackage{graphicx}
\usepackage{geometry}
\PassOptionsToPackage{normalem}{ulem}
\usepackage{ulem}
\usepackage[pdfusetitle,
 bookmarks=true,bookmarksnumbered=true,bookmarksopen=false,
 breaklinks=false,pdfborder={0 0 1},backref=false,colorlinks=false]
 {hyperref}

\makeatletter

\makesavenoteenv{tabular}

\providecommand{\tabularnewline}{\\}
\newenvironment{cellvarwidth}[1][t]
    {\begin{varwidth}[#1]{\linewidth}}
    {\@finalstrut\@arstrutbox\end{varwidth}}
\providecolor{lyxadded}{rgb}{0,0,1}
\providecolor{lyxdeleted}{rgb}{1,0,0}
\DeclareRobustCommand{\mklyxadded}[1]{\textcolor{lyxadded}\bgroup#1\egroup}
\DeclareRobustCommand{\mklyxdeleted}[1]{\textcolor{lyxdeleted}\bgroup\mklyxsout{#1}\egroup}
\DeclareRobustCommand{\mklyxsout}[1]{\ifx\\#1\else\sout{#1}\fi}

\newcommand{\lyxaddress}[1]{
	\par {\raggedright #1
	\vspace{1.4em}
	\noindent\par}
}

\ifdefined\showcaptionsetup
 \PassOptionsToPackage{caption=false}{subfig}
\fi
\usepackage{subfig}
\makeatother

\begin{document}
\title{A conformally-Euclidean Line Element for evaluating small color differences}
\author{Patrick De Visschere$^{1}$, Patrick Candry$^{1},$ Kristiaan Neyts$^{2}$}
\maketitle

\lyxaddress{$^{1}$Ghent University, Department Electronics and Information Systems,
Liquid Crystals and Photonics.\\
Technologiepark-Zwijnaarde 126, BE-9052 Gent, email: Patrick.DeVisschere@UGent.be,
ORCID: \url{https://orcid.org/0000-0003-0278-8199}}

\lyxaddress{$2$Hong Kong University of Science and Technology, Department of
Electronic \& Computer Engineering, Clear Water Bay Road, Hong Kong
SAR.}
\begin{abstract}
Starting from our previously proposed physiologically based line element
(LE) and considering more ``surface color'' datasets, we derive
a simplified version which matches colorimeter and display datasets
equally well (based on an F-test) and surface datasets significantly
better. It resulted into a conformally-Euclidean line element, which
is conceptually much simpler than any existing color difference metrics.
The color difference is written as an Euclidean difference multiplied
with a simple factor which depends on the luminance only. In a subspace
with constant luminance, as considered by MacAdam, this factor becomes
constant and the subspace is flat. The same holds for sufficiently
large luminances. Based on this LE we derive perceptual coordinates
$\left(A,l_{c},s_{c}\right)$. These coordinates are completely novel,
are thus physiologically based and are very similar to the CIELab
$\left(L^{*},a^{*},b^{*}\right)$ coordinates.
\end{abstract}

\section{Introduction}

In 1976 the CIE introduced, besides the $L^{*}u^{*}v^{*}$ space,
the $L^{*}a^{*}b^{*}$ space \cite{Robertson:1977oi}, both being
intended as approximately uniform color spaces, meaning that color
differences could approximately be derived as Euclidean distances
in these spaces
\begin{equation}
{\Delta E^{*}_{ab}}^{2}=\left(\Delta L^{*}\right)^{2}+\left(\Delta a^{*}\right)^{2}+\left(\Delta b^{*}\right)^{2}\label{eq:Lab}
\end{equation}

Today it is widely accepted that the human visual color space is not
Euclidean and Eq.(\ref{eq:Lab}) is indeed at best an approximation.
Nevertheless these $L^{*},a^{*},b^{*}$ values are still legitimate
as local coordinates and in the intervening time many attempts have
been made to define a more accurate line element (LE), expressed more
generally by a suitable 3-dimensional symmetric tensor
\begin{equation}
{d\sigma}^{2}=\begin{bmatrix}dL^{*} & da^{*} & db^{*}\end{bmatrix}\cdot\left[g_{ij}\right]\cdot\begin{bmatrix}dL^{*}\\
da^{*}\\
db^{*}
\end{bmatrix}
\end{equation}

where the tensor elements $g_{ij}=g_{ji}$ vary with the color point
with coordinates $L^{*},a^{*},b^{*}$. Such a space is known as a
Riemannian manifold supposing the tensor is positive definite. The
best result obtained so far along these lines resulted into the CIEDE2000
color difference formula \cite{Luo:2001hd}, and its descendants like
CIECAM02 \cite{Luo2013}. CIEDE2000 uses slightly different coordinates
$L^{\prime},a^{\prime},b^{\prime}$, switches to polar coordinates
with chroma $C^{\prime}=\sqrt{a^{\prime2}+b^{\prime2}}$ and hue angle
$h^{\prime}=\arctan(b^{\prime}/a^{\prime})$ and besides the diagonal
elements of $g$, $S^{-2}_{L},S^{-2}_{C},S^{-2}_{H}$, introduces
a chromatic non-diagonal element involving a rotation term $R_{T}$.
Without this rotation term the chromatic threshold ellipses are aligned
with their long axis along the radial direction and $R_{T}$ introduces
a rotation of these ellipses away from the radial direction in a very
limited part of the ``blue'' chromatic region. Melgosa et al. \cite{Melgosa:2004wl}
found that of all these improvements of CIEDE2000 compared with previous
models, the transformation from $a^{*}\rightarrow a^{\prime}$ has
the least effect. We will show that its effect is actually negative
(see Table~\ref{tab:F-tests-CIEDE2000}).

In \cite{Candry:2022kx} we presented an alternative line element
based on earlier attempts by Friele \cite{Friele:1978fy} and compared
with the more empirical CIE models was build on more physiologically
solid ground. The organization of the human visual system from the
retina over the LGN (Lateral Geniculate Nucleus) and then further
down different layers in the visual cortex has been thoroughly studied
and certainly the initial stages are now well understood. It seemed
logical to implement that knowledge as much as possible in a model
for the prediction of visually perceived color differences. In particular
we have chosen physiological coordinates (MacLeod-Boynton coordinates
\cite{MacLeod:1979pa,Boynton:1996lo}), this in contrast with the
CIE color difference metric which uses perceptual coordinates. We
believe that perceptual coordinates will emerge automatically from
a sound physiological model. Our LE was primarily evaluated against
color difference data obtained with colorimeters, including MacAdam's
famous results \cite{MacAdam:1942kk}\cite{Brown:1949fp} but performance
against the RIT-Dupont dataset \cite{Berns:2009jv} obtained with
surface colors was also included. To evaluate a model against an experimental
dataset we used a newly developed difference measure $d$, based on
the eigenvalues of the metric tensors $\left[g_{ij}\right]$ to be
compared. We showed that the many color difference measures which
have been proposed ($V_{AB}$, CV, $\gamma$, STRESS, R) and our eigenvalue
based $d$ are in fact all equivalent for small color differences
and that for larger color differences they are also very similar \cite{Visschere:2026mw}.

Since our initial publication \cite{Candry:2022kx} we have tested
our LE against many more datasets \cite{Candry:2025ax}, primarily
obtained for surface colors but also for experiments performed with
displays. We found that the CIEDE2000 LE does not perform as well
for colorimeter data and for display data, while our LE performed
still reasonably well for surface colors, although not as well as
for colorimeters and displays, but clearly not as well as CIEDE2000.
Since we find it unlikely that the human visual system would sustain
two different systems for evaluating color differences and our LE
was better grounded physiologically we believed it should be possible
to modify our LE so that it could be turned into a generally valid
LE. Analysing the geometrical properties of our LE we eventually found
that it could be turned into a conformally-Euclidean LE, which is
perhaps the next simple LE after an Euclidean one and defined by \cite{Lee:2018lj}

\begin{equation}
{d\sigma}^{2}=e^{2f}\left[\left(dx^{1}\right)^{2}+\left(dx^{2}\right)^{2}+\left(dx^{3}\right)^{2}\right]\label{eq:aconformalLE}
\end{equation}

where $d\sigma$ is an infinitesimal (just noticeable) color difference
(JND) and $f$ is a scalar function of the (perceptual) coordinates
$x^{1},x^{2},x^{3}$ and the exponential guarantees that this multiplication
factor is positive. As will become clear the coordinates $x^{i}$
we have found are very similar with $L^{*},a^{*},b^{*}$, but with
some marked differences. With hindsight we can then conclude that
what was needed in (\ref{eq:Lab}) was not a much more complex tensor
but maybe slightly different coordinates.

We will refer to our original LE as RieLE1 ~(This is an abbreviation
of Riemannian LE, but refers also to Friele.) \cite{Candry:2022kx}
and to the conformal one as RieLE2. In section~\ref{sec:The-development-of}
we define RieLE2 emphasizing the differences with RieLE1. In section~\ref{sec:The-performance-of}
the predictive performance of RieLE2 is compared with those of CIEDE2000
and RieLE1 and in section~\ref{sec:Perceptual-coordinates-for} we
derive the perceptual coordinates leading to Eq.(\ref{eq:aconformalLE}).
The geometric properties (geodesics, curvature, ...) of the resulting
new color space will be presented in a separate paper.

\section{The definition of RieLE2}\label{sec:The-development-of}

RieLE2 and RieLE1 \cite{Candry:2022kx} are defined using the retinal
illuminance (actually $\ln Y$) and the MacLeod-Boynton coordinates
$l=L/Y$ and $s=S/Y$ as coordinates, where $Y=L+M$ and $L,M,S$
are based on the fundamental cone responses~ Originally we used the
Smith-Pokorny cone fundamentals but here we switched to the more recent
Stockman-Sharpe fundamentals \cite{Stockman:2000kq}. The main thresholds
of RieLE2 are defined by a diagonal tensor
\begin{equation}
g_{0}=\begin{bmatrix}g_{11} & 0 & 0\\
0 & g_{22} & 0\\
0 & 0 & g_{33}
\end{bmatrix}\label{eq:g0}
\end{equation}

where 
\begin{equation}
g_{11}=\frac{1}{\psi^{2}_{A}}\quad g_{22}=\frac{1}{\psi^{2}_{T}}\quad g_{33}=\frac{1}{\psi^{2}_{D}}
\end{equation}
and $\psi_{A/T/D}$ are the thresholds of the achromatic, the red-green
and the yellow-blue channels which are defined by
\begin{equation}
\psi_{A}=f_{A}k_{0}\label{eq:fAdef}
\end{equation}
\begin{equation}
\psi_{T}=f_{c}\tilde{\psi}_{T}\label{eq:psiTc-1}
\end{equation}
\begin{equation}
\psi_{D}=f_{c}\tilde{\psi}_{D}\label{eq:psiDc-1}
\end{equation}

with
\begin{equation}
\tilde{\psi}_{T}=k_{1}l_{E}+k_{2}\left|l-l_{a}\right|\label{eq:psiTtilde}
\end{equation}

and
\begin{equation}
\tilde{\psi}_{D}=k_{3}s_{E}+k_{4}\left|s-s_{a}\right|\label{eq:psiDtilde}
\end{equation}

and where 
\begin{equation}
f_{A}=\sqrt{1+\frac{Y_{A}}{Y}}\quad f_{c}=\sqrt{1+\frac{Y_{c}}{Y}}\label{eq:fcdef-1}
\end{equation}

and $Y_{A},Y_{c},k_{0..4}$ are constants, $\left(l_{a},s_{a}\right)$
are the $\left(l,s\right)$-values of the adapting color, and $\left(l_{E},s_{E}\right)$
those of equal energy white. The factors $f_{A},f_{c}$ implement
the transition from de Vries-Rose to Weber behavior for the achromatic
channel ($f_{A}$) and for the chromatic channels ($f_{c}$).

The complete metric tensor of RieLE2 $g$ can be written succinctly
as the product of this diagonal tensor $g_{0}$ and a ``rotation''
tensor
\begin{equation}
\Delta=\begin{bmatrix}1 & \delta_{12} & 0\\
\delta_{12} & 1+\delta^{2}_{12} & \delta_{23}\\
0 & \delta_{23} & 1+\delta^{2}_{23}
\end{bmatrix}\label{eq:delta1}
\end{equation}

with
\begin{equation}
g=\sqrt{g_{0}}\cdot\Delta\cdot\sqrt{g_{0}}\label{eq:g}
\end{equation}

and the LE
\begin{equation}
d\sigma^{2}=\begin{bmatrix}\frac{dY}{Y} & dl & ds\end{bmatrix}\cdot\left[g_{ij}\right]\cdot\begin{bmatrix}\frac{dY}{Y}\\
dl\\
ds
\end{bmatrix}\label{eq:LE}
\end{equation}

The functions $\delta_{12}$ and $\delta_{23}$ model the rotation
of the ellipsoid away from the main axes in respectively the $\left(dY/Y,dl\right)$-plane
and in the chromatic $\left(dl,ds\right)$-plane. The determinant
of $g$ determines the volume of the ellipsoid and is given by $\det g=\det g_{0}$
and is independent of the rotation terms, which are given by
\begin{equation}
\delta_{12}=k_{12}\left(l-l_{a}\right)^{2}\label{eq:delta12}
\end{equation}
\begin{equation}
\delta_{23}=k_{23}\label{eq:delta23_1}
\end{equation}

We summarize the main differences with RieLE1. (1) In (\ref{eq:delta1})
the squares $\delta^{2}_{12}$ and $\delta^{2}_{23}$ have been added
to the chromatic diagonal terms. This ensures the independence of
$\det g$ from the rotation terms and guarantees that for $Y\rightarrow\infty$
the curvature tends to zero and the LE thus becomes Euclidean. (2)
$\delta_{23}$ was a function of $s$ in RieLE1 and is replaced by
a constant $k_{23}$ in RieLE2, which improves the matching of surface
color datasets. (3) The $Y$-dependencies of the chromatic thresholds
$\psi_{T/D}$ have been modified in a way that the LE becomes conformally
Euclidean. (4) Some additional minor simplifications have been applied
to the thresholds $\psi_{A/T/D}$.

In section~\ref{sec:Perceptual-coordinates-for} we will show that
apart from the factor $f_{c}$ this LE is Euclidean with the conformal
factor $e^{2f}=f^{-2}_{c}$. For convenience we define a reduced metric
tensor $\tilde{g}$ not including this conformal factor
\begin{equation}
g=\frac{1}{f^{2}_{c}}\tilde{g}
\end{equation}

This can still be written as
\[
\tilde{g}=\sqrt{\tilde{g}_{0}}\cdot\Delta\cdot\sqrt{\tilde{g}_{0}}
\]

with
\[
\tilde{g}_{0}=\begin{bmatrix}\tilde{g}_{11} & 0 & 0\\
0 & \tilde{g}_{22} & 0\\
0 & 0 & \tilde{g}_{33}
\end{bmatrix}
\]
\[
\tilde{g}_{11}=\frac{1}{\tilde{\psi}^{2}_{A}}\quad\tilde{g}_{22}=\frac{1}{\tilde{\psi}^{2}_{T}}\quad\tilde{g}_{33}=\frac{1}{\tilde{\psi}^{2}_{D}}
\]

where the chromatic thresholds have been defined in (\ref{eq:psiTtilde})(\ref{eq:psiDtilde})
and 
\[
\tilde{\psi}_{A}=k_{0}\frac{f_{A}}{f_{c}}
\]

is still a function of $Y$ only whereas $\tilde{\psi}_{T}$ depends
on $l$ only and $\tilde{\psi}_{D}$ similarly on $s$ only.

Using RieLE1 we previously estimated the number of discernible colors
\cite{Candry:2023ix} and we published a version applicable to sinusoidal
color variations (gratings) \cite{Candry:2024kx}. Adapting these
calculations to RieLE2 is straightforward. Since the grating experiments
are done on displays and with the observer adapted to the test color
we expect no significant differences. Similarly the number of discernible
colors was primarily meant for displays and we don't expect important
changes neither.

\section{The performance of RieLE2}\label{sec:The-performance-of}

RieLE1 \cite{Candry:2022kx} was mainly developed against color matching
data obtained with colorimeters although also a surface color set
was included (RIT-Dupont, code rd90 \cite{Berns:2009jv}). RieLE2
was tested against many more datasets \cite{Candry:2025ax}, in particular
many more surface color datasets. We also included datasets obtained
with display monitors. These datasets and their main parameters are
listed in Appendix~\ref{sec:Overview-of-the}. RieLE2 depends, besides
on $Y_{A/c}$ on 9 parameters: $k_{0-4}$, $k_{12}$ and $k_{23}$,
where for $k_{2\pm}$ and $k_{4\pm}$ separate values are chosen for
the positive/negative excursions of $l,s$ w.r.t. $l_{a},s_{a}$,
since we believe that these positive and negative pathways are handled
by separate neurologic networks \cite{Sankeralli:2001nd}. These parameters
are chosen optimally for each dataset, but we also included the results
using group values (for colorimeters, displays, surface colors and
for experiments with a dark surround separately) and also using generic
parameters for aperture colors \{colorimeter, display, dark surround\}
and surface colors.

In all cases the performance of RieLE2 is obtained by calculating
the rms value $d_{\mathrm{rms}}=\sqrt{\left(\sum_{i}d^{2}_{i}\right)/N}$
of the $N$ individual $d_{i}$ values where all ellipsoids of a given
set are uniformly scaled so as to make $d_{\mathrm{rms}}$ minimal.
Parameters are extracted setting the size $d\sigma$ occurring in
Eq.(\ref{eq:aconformalLE}) equal to one. As experiments to evaluate
if colors are distinguishable are made with different procedures (which
questions are asked, one or two eyes, illumination condition...),
the size of the ellipsoids should be adjusted for every dataset separately.
This is done by a scaling procedure outlined in Appendix~\ref{sec:Scaling-of-the}.
The predictions of CIEDE2000 were analyzed in the same way but using
the generic parameter values with the so called parametric factors
$k_{L}$, $k_{C}$ and $k_{H}$ set to one and with these parametric
factors optimized for each dataset.

To obtain the parameters $k_{i,\mathrm{group}}$ for a group of datasets
(e.g. the colorimeter datasets) all ellipsoids were optimized together
in the same way but where a different scaling factor was applied to
the ellipsoids of each member dataset. A few datasets have been obtained
monocularly. We have divided the ellipsoid scale sensitive parameters
by $\sqrt{2}$ to align them with the remaining binocular ones. The
found parameters were then rounded to two significant digits. The
group parameters are listed in Table~\ref{tab:parameter-values}
and the corresponding scale factors are shown in the tables in Appendix~\ref{sec:Overview-of-the}
under the heading ``$d\sigma_{\mathrm{group}}$''. E.g. for the
BFD-P dataset the ellipses are on average sized with $d\sigma_{\mathrm{group}}=5.7$.
Our LE predicts JNDs and we make no difference between threshold and
supra threshold color differences. For larger color differences the
LE should be integrated along the geodesic between the color points
considered.

The matching performance is only weakly dependent on the choice of
$Y_{A/c}$. We believe this is due to the lack of data for different
(low) luminance luminance levels. Therefore we have made a fixed plausible
choice based partly on available evidence and partly for simplicity.
Rovamo et al. found with isoluminant sinusoidal gratings a value $Y_{c}=165\,td$
for the chromatic critical luminance, marking the transition from
de Vries-Rose to Weber behavior, independent of frequency, and with
luminance gratings a frequency dependent $Y_{A}=11.6u^{2}$ \cite{Rovamo:2001ye}.
All the experiments considered in this work have been obtained with
a split field set-up where the reference color and the test color
are presented side-by-side. With the edge in our view being dominant
we believe the effective $Y_{A}$ being rather large and we set $Y_{A}=800\,td$,
corresponding with a frequency of nearly 8~cpd. When $Y_{A}\neq Y_{c}$
the base lightness function $\tilde{L}\left(Y\right)$ becomes rather
complicated (see Appendix~\ref{sec:The-lightness-functions}) but
the numerical constants emerging in these expressions are somewhat
simpler for $Y_{A}=4Y_{c}$, hence we settled for $Y_{A}=800\,td$
and $Y_{c}=200\,td$.

\begin{figure}
\subfloat[Colorimeters]{\begin{centering}
\includegraphics[width=0.45\columnwidth]{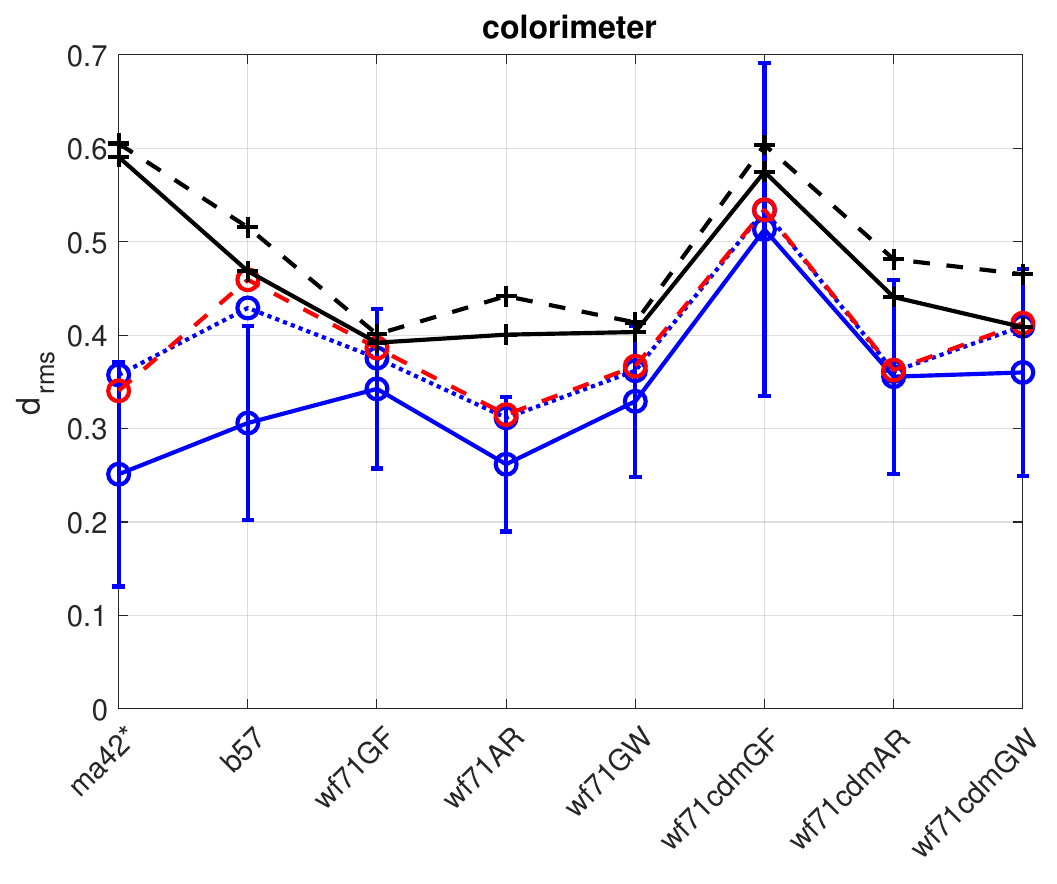}
\par\end{centering}
}\hfill{}\subfloat[Displays]{\begin{centering}
\includegraphics[width=0.45\columnwidth]{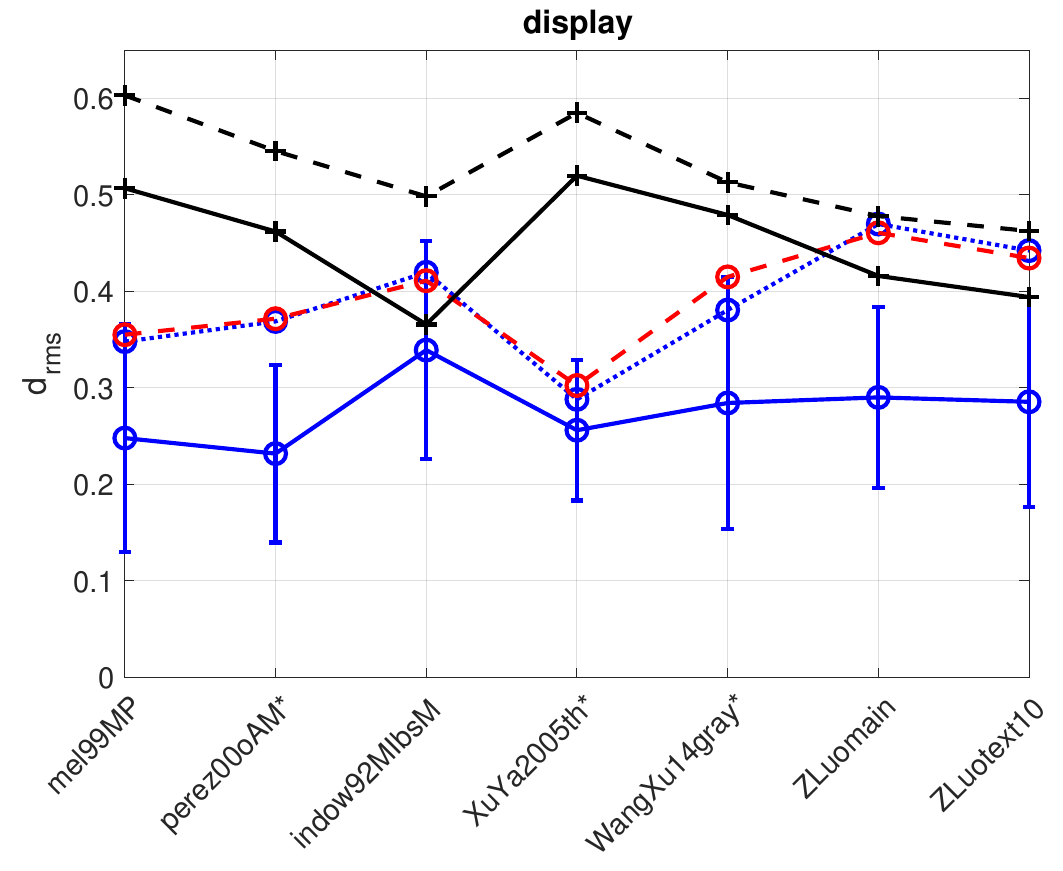}
\par\end{centering}
}

\subfloat[Surface colors]{\begin{centering}
\includegraphics[width=0.45\columnwidth]{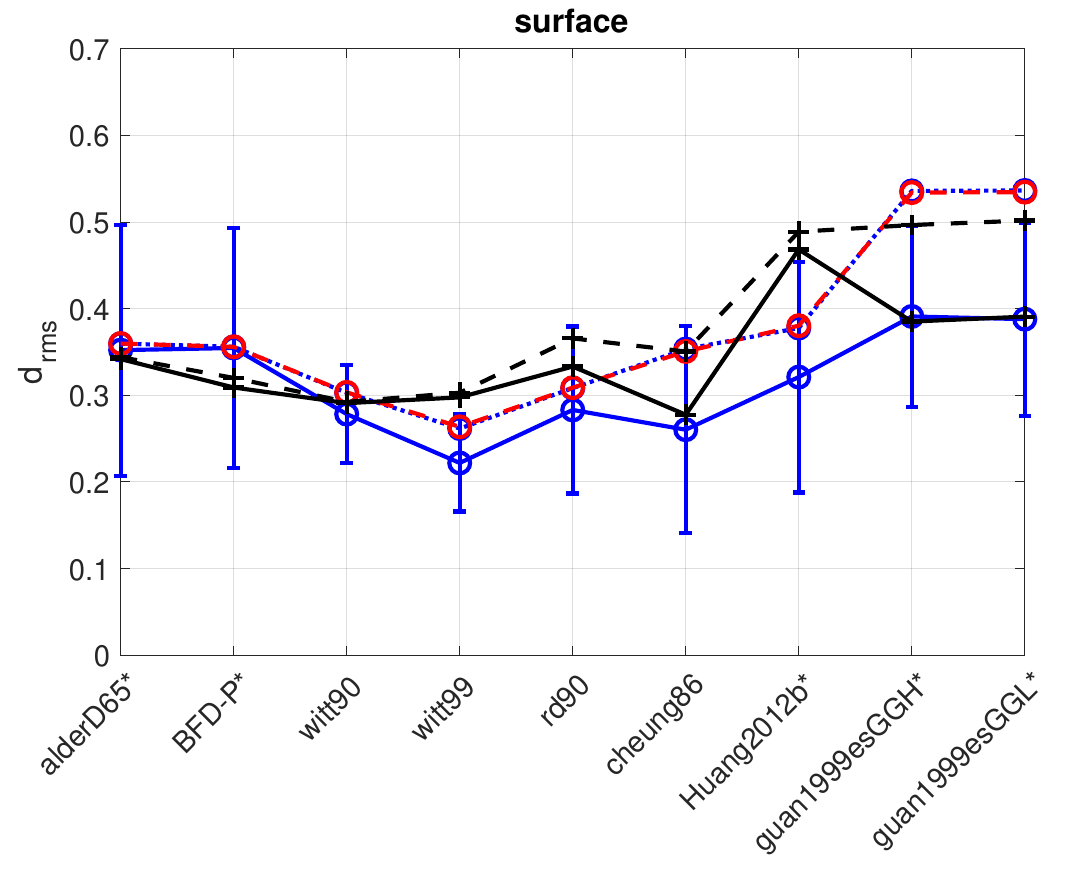}
\par\end{centering}
}\hfill{}\subfloat[BFD-P surface colors]{\begin{centering}
\includegraphics[width=0.45\columnwidth]{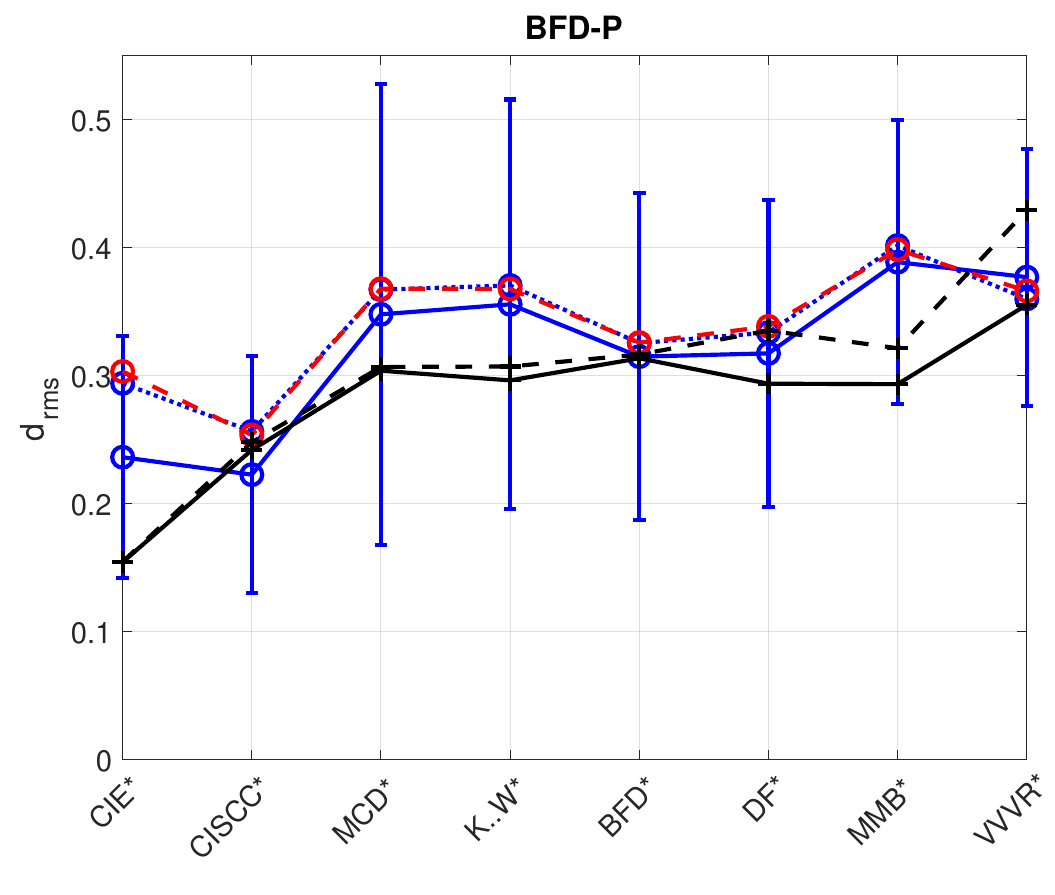}
\par\end{centering}
}

\subfloat[Dark surround (colorimeters, displays).]{\begin{centering}
\includegraphics[width=0.45\columnwidth]{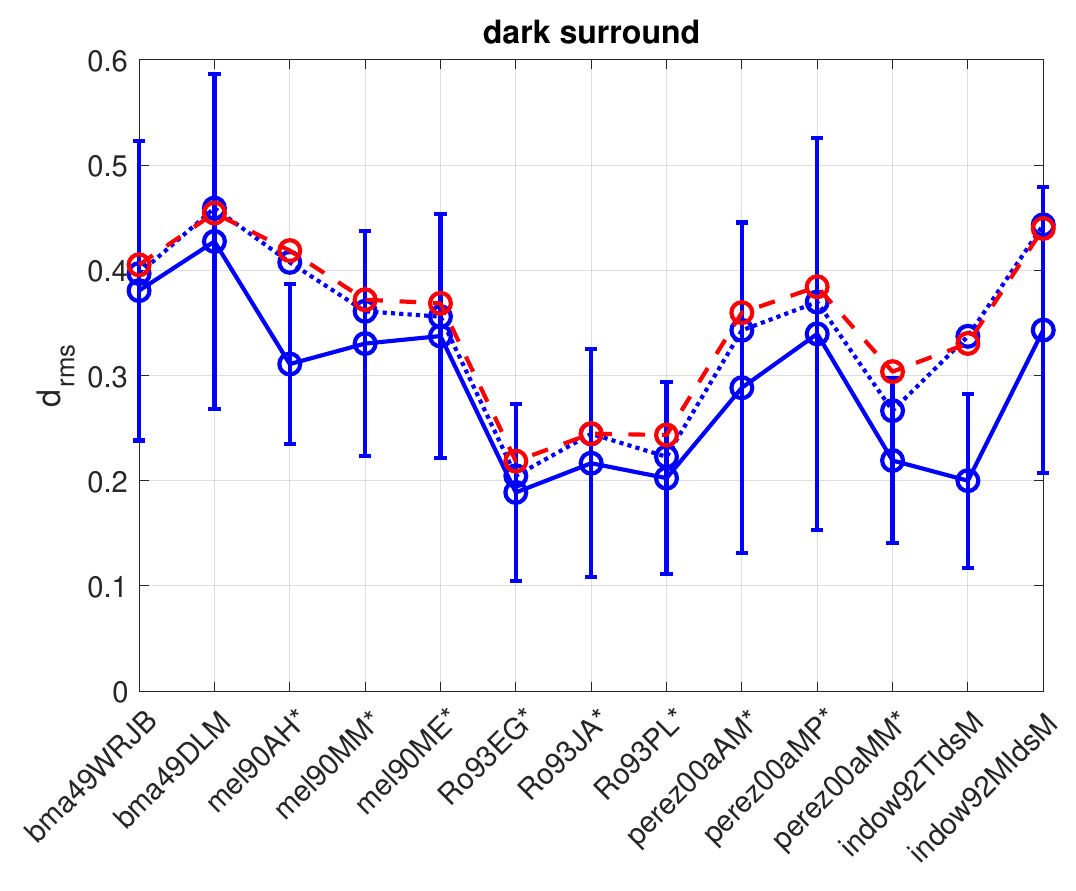}
\par\end{centering}
}

\caption{Overview of the eigenvalue based $d_{rms}$ measure for optimal dataset,
group and generic parameters for different types of medium. The $d_{rms}$
obtained for optimal parameters per dataset are connected by full
lines in blue, the ones obtained for group level parameters by dotted
lines in blue and for global parameters with dashed lines in red.
The CIEDE2000 predictions (marked by crosses) with $k_{L}=k_{C}=k_{H}=1$
are shown in black and connected by dashed lines and for optimised
$k_{L},k_{C},k_{H}$ in full lines The errorbars show the standard
deviation for the optimal $d_{rms}$ . Datasets marked with a {*}
are \textquotedblleft chromatic only\textquotedblright , meaning they
measured the $Y$-section ellipse.}\label{fig:results}
\end{figure}

In Fig.~\ref{fig:results} the $d_{\mathrm{rms}}$ are shown for
optimal parameters $k_{i,\mathrm{set}}$ and for parameters $k_{i,\mathrm{group}}$
optimized per group for respectively experiments done with colorimeters,
displays, surface colors and for experiments done with a dark surround.
As a comparison the result obtained with the CIEDE2000 expression
is also shown ( for $k_{L}=k_{C}=k_{H}=1$ and for dataset optimized
$k_{L},k_{C},k_{H}$). For each group we have also averaged the performances
taking into account the number of color points in each dataset. These
numbers are shown in Table~\ref{tab:rms-averaged-value} together
with their minimal and maximal values. Since $d^{2}_{\mathrm{rms}}$
is merely a variance it can be used to perform an F-test (significant
level tested 5\%) revealing whether two models are significantly different
or not as shown in Table~\ref{tab:F-tests-CIEDE2000} and Table~\ref{tab:F-tests-RieLE}.
For the ``colorimeter'' and ``display'' datasets RieLE2 is significantly
better than CIEDE2000. For the ``surface'' datasets CIEDE2000 is
not significantly better. We have included also the results for CIEDE2000
without the $a^{*}\rightarrow a^{\prime}$ transform included ($G=0$)
and in that case CIEDE2000 is significantly better than RieLE2 for
the BFD-P set but not for the wider ``surface'' set and RieLE2 becomes
only marginally better for the ``display'' datasets. RieLE2 is significantly
better than RieLE1 for the ``surface'' datasets and there is no
significant difference for the other media.

\begin{figure}
\begin{centering}
\subfloat[]{\includegraphics[width=7cm]{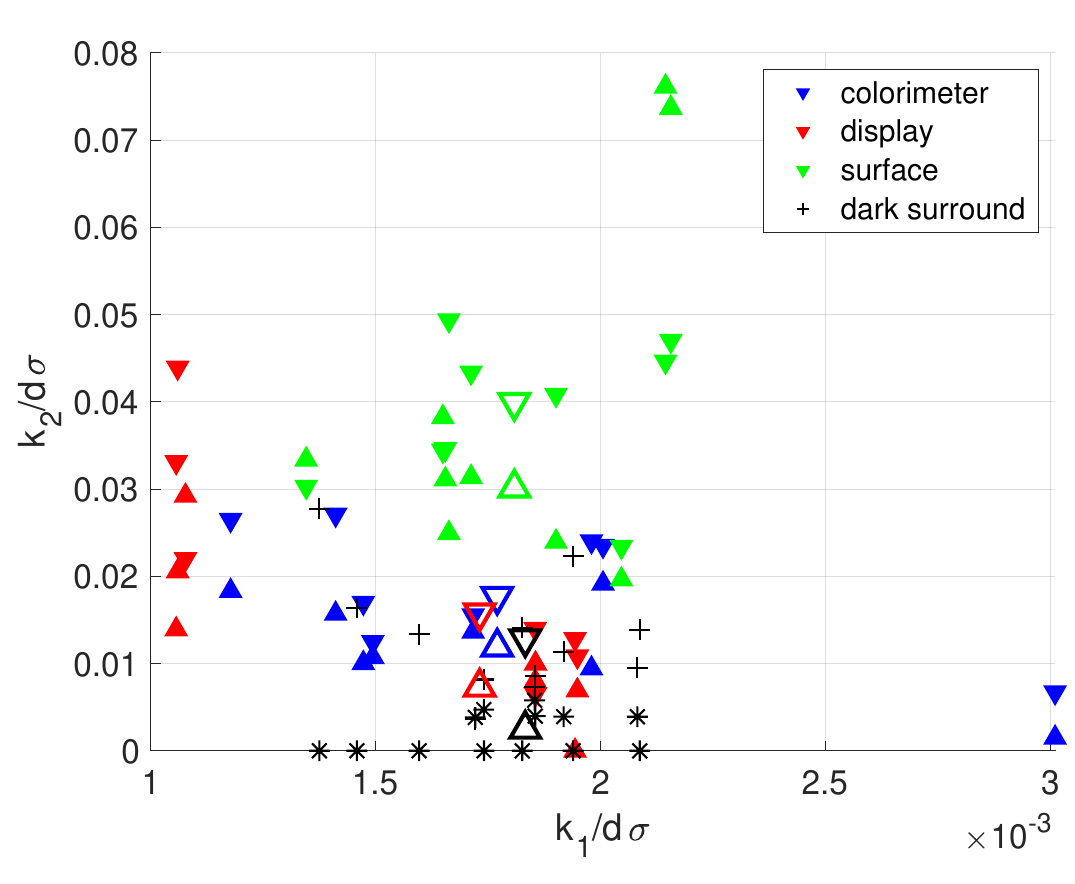}

}\hspace{1cm}\subfloat[]{\includegraphics[width=7cm]{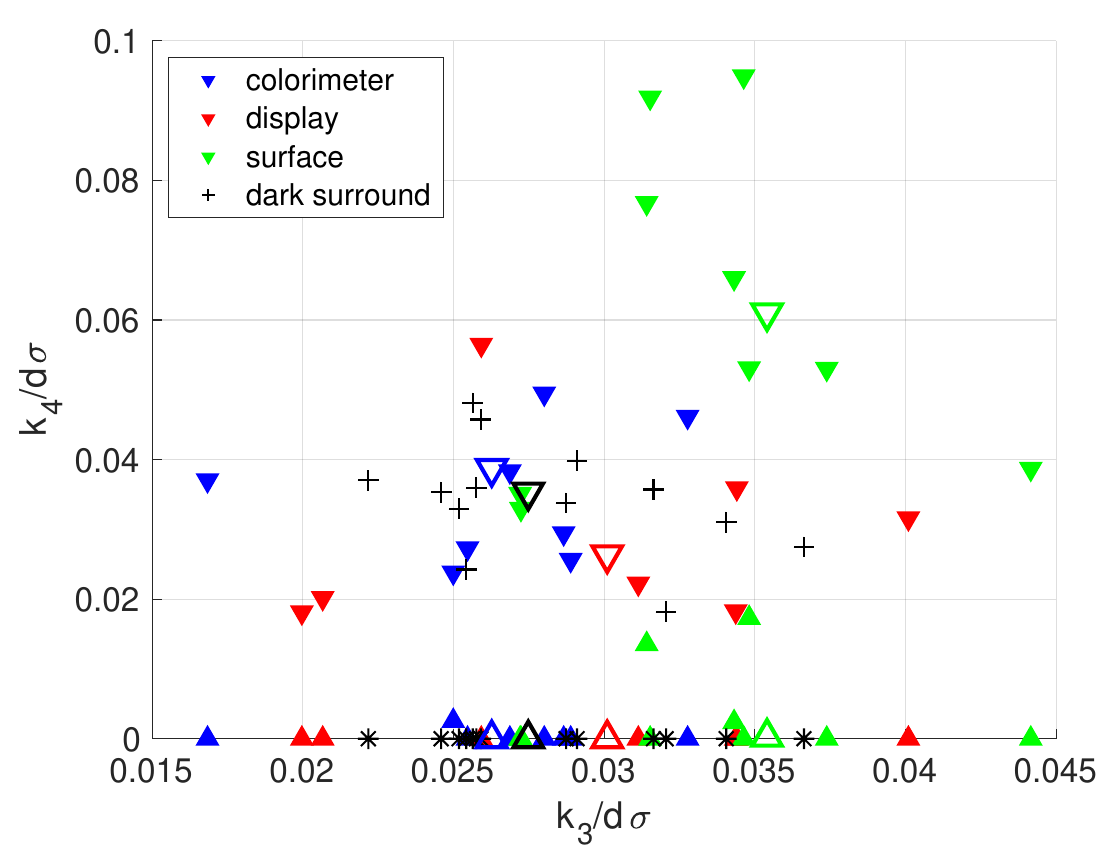}

}
\par\end{centering}
\begin{centering}
\subfloat[]{\includegraphics[width=7cm]{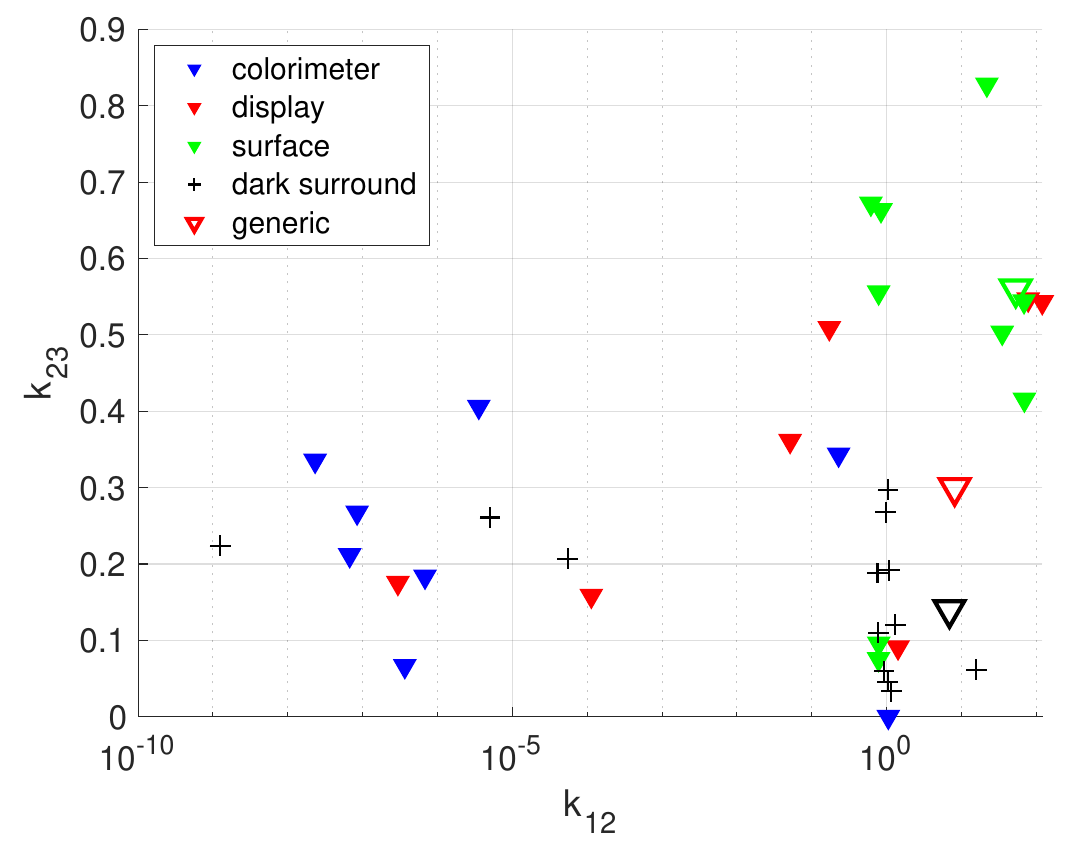}

}
\par\end{centering}
\caption{Plot of the optimal parameters $k_{\mathrm{set}}$ for the datasets
in the 4 groups, normalized with their sizes $d\sigma$ (a) $k_{2}/d\sigma$
versus $k_{1}/d\sigma$, (b) $k_{4}/d\sigma$ versus $k_{3}/d\sigma$
and the scale invariant parameters $k_{23}$ versus $k_{12}$. The
parameters $k_{2+}$ and $k_{4+}$ for positive excursion w.r.t. the
adapting point are shown with a down pointing triangle and those for
negative excursion with an up pointing one. The group level parameter
values $k_{\mathrm{group}}$ scaled with the scale factors $F_{g}$
and $F$ are shown as open triangles. For the aperture groups \textquotedblleft colorimeter\textquotedblright ,
\textquotedblright display\textquotedblright{} and \textquotedblleft dark
surround\textquotedblright{} they have been replaced by the generic
$k_{\mathrm{gen}}$ . As explained in Appendix~\ref{sec:Scaling-of-the}
$k_{\mathrm{set}}/d\sigma\approx k_{\mathrm{group}}F_{g}F\approx k_{\mathrm{gen}}$.}\label{fig:kplots}
\end{figure}

The best values obtained are for the optimal dataset parameters because
in this case parameters are optimized for the individual datasets,
and they are in the range $d_{\mathrm{rms}}=0.19-0.25$ with averages
as low as 0.27 for ``displays'' and varying between 0.32-0.37 for
the other groups. The worst values are again lowest for ``displays''
(0.34) and in the range 0.39-0.51, where the latter is for ``colorimeters''
and looks like an exception (one of three observers of dataset wf71cdm
performing significantly worse). The deviations obtained with the
group parameters are higher but modest increases hold for the ``surface''
group and in particular the ``BFD-P'' group. These datasets were
streamlined by Luo \cite{Luo:1986fp} and among other things they
were scaled for a fixed luminance level, but we used the original
values, since $Y$-dependence turns out as the sole parameter influencing
the curvature of the color space. Replacing the group parameters by
generic ones has a very small penalty. The parameters $k_{i,\mathrm{set}}$,
$k_{i,\mathrm{group}}$ and $k_{i,\mathrm{gen}}$ are shown in Fig.\ref{fig:kplots}.
The optimal values for $k_{4-}$ are zero for all groups. Those for
$k_{2-}$ are only zero for the ``dark surround'' group. For the
aperture groups ``colorimeter'', ``display'' and ``dark surround'',
the $k_{12}$ values fall apart into a cluster $\approx0$ and a cluster
averaging around $k_{12}\approx1$. The surface group also clusters
with the latter group but shows higher $k_{23}$ values. Fig.\ref{fig:kplots}(a)
shows 3 outliers, two from the ``surface'' group (the guan1999 sets)
and one from the ``colorimeter'' group (the b57 set). A comparison
between the predictions of the CIEDE2000 model and RieLE2 for the
generic surface parameters is given in Supplement~1.

This section provides a line element for estimating the shape and
size of JND ellipsoids around arbitrary color points. It seems that
each class of experiments (displays, surface colors, ...) has its
own set of parameters to determine the JND. There is a satisfactory
agreement with a large number of datasets, even when the group parameters
are used. The JND ellipsoids can be represented in any color coordinate
system, when the appropriate transformation is implemented. In the
next section, we propose a new color coordinate system, in which the
JND ellipses reduce to spheres, with the radius of the spheres depending
on the value of $Y$ only.

\begin{table}
\caption{rms averaged value of $d_{\mathrm{rms}}$ for each group of datasets
for dataset optimized, for group optimized and for generic parameter
values, for CIEDE2000, with standard parameters and with dataset optimized
$k_{L},k_{C},k_{H}$. In the average the number of color points in
the datasets have been taken into account. Besides the averages the
minimum and maximum values are also mentioned between brackets.}\label{tab:rms-averaged-value}

\centering{}%
\begin{tabular}{|c|c|c|c|c|c|}
\hline 
\multirow{2}{*}{group} & \multicolumn{3}{c|}{RieLE2} & \multicolumn{2}{c|}{CIEDE2000}\tabularnewline
\cline{2-6}
 & \begin{cellvarwidth}[t]
\centering
optimal \\
dataset
\end{cellvarwidth} & \begin{cellvarwidth}[t]
\centering
group\\
$k_{\mathrm{group}},d\sigma_{\mathrm{group}}$
\end{cellvarwidth} & \begin{cellvarwidth}[t]
\centering
generic \\
$k_{\mathrm{gen}},d\sigma$
\end{cellvarwidth} & \begin{cellvarwidth}[t]
\centering
standard\\
1:1:1
\end{cellvarwidth} & \begin{cellvarwidth}[t]
\centering
optimal\\
$k_{L},k_{C},k_{H}$
\end{cellvarwidth}\tabularnewline
\hline 
\hline 
colorimeter & \begin{cellvarwidth}[t]
\centering
0.36\\
(0.25-0.51)
\end{cellvarwidth} & \begin{cellvarwidth}[t]
\centering
0.41\\
(0.31-0.54)
\end{cellvarwidth} & \begin{cellvarwidth}[t]
\centering
0.41\\
(0.31-0.53)
\end{cellvarwidth} & \begin{cellvarwidth}[t]
\centering
0.50\\
(0.40-0.61)
\end{cellvarwidth} & \begin{cellvarwidth}[t]
\centering
0.46\\
(0.39-0.59)
\end{cellvarwidth}\tabularnewline
\hline 
display & \begin{cellvarwidth}[t]
\centering
0.27\\
(0.23-0.34)
\end{cellvarwidth} & \begin{cellvarwidth}[t]
\centering
0.40\\
(0.30-0.49)
\end{cellvarwidth} & \begin{cellvarwidth}[t]
\centering
0.39\\
(0.30-0.46)
\end{cellvarwidth} & \begin{cellvarwidth}[t]
\centering
0.54\\
(0.46-0.60)
\end{cellvarwidth} & \begin{cellvarwidth}[t]
\centering
0.46\\
(0.37-0.52)
\end{cellvarwidth}\tabularnewline
\hline 
dark surround & \begin{cellvarwidth}[t]
\centering
0.32\\
(0.19-0.43)
\end{cellvarwidth} & \begin{cellvarwidth}[t]
\centering
0.40\\
(0.25-0.53)
\end{cellvarwidth} & \begin{cellvarwidth}[t]
\centering
0.36\\
(0.22-0.45)
\end{cellvarwidth} & \multicolumn{2}{c|}{n.a.}\tabularnewline
\hline 
\hline 
surface & \begin{cellvarwidth}[t]
\centering
0.34\\
(0.22-0.39)
\end{cellvarwidth} & \begin{cellvarwidth}[t]
\centering
0.36\\
(0.26-0.54)
\end{cellvarwidth} & \begin{cellvarwidth}[t]
\centering
0.36\\
(0.25-0.53)
\end{cellvarwidth} & \begin{cellvarwidth}[t]
\centering
0.35\\
(0.29-0.50)
\end{cellvarwidth} & \begin{cellvarwidth}[t]
\centering
0.33\\
(0.28-0.47)
\end{cellvarwidth}\tabularnewline
\hline 
BFD-P & \begin{cellvarwidth}[t]
\centering
0.33\\
(0.22-0.39)
\end{cellvarwidth} & \begin{cellvarwidth}[t]
\centering
0.35\\
(0.26-0.40)
\end{cellvarwidth} & \begin{cellvarwidth}[t]
\centering
0.35\\
(0.25-0.40)
\end{cellvarwidth} & \begin{cellvarwidth}[t]
\centering
0.32\\
(0.15-0.43)
\end{cellvarwidth} & \begin{cellvarwidth}[t]
\centering
0.30\\
(0.15-0.36)
\end{cellvarwidth}\tabularnewline
\hline 
\end{tabular}
\end{table}

\begin{table}
\centering{}\caption{F-tests for $d^{2}_{\mathrm{rms}}$between RieLE2 (generic) and CIEDE2000,
without the $a^{*}\rightarrow a^{\prime}$ transform ($G=0$) and
for the standard with this transform ($G\protect\neq0$), revealing
that in all these cases the performance is less in the latter case.
$F_{\mathrm{crit}}=F^{-1}\left(0.975,N-1,N-1\right)$.}\label{tab:F-tests-CIEDE2000}
\begin{tabular}{|c|c|c|c|c|c|c|c|c|}
\hline 
\multirow{2}{*}{group} & \multirow{2}{*}{$N$} & \multirow{2}{*}{RieLE2} & \multicolumn{4}{c|}{CIEDE2000} & \multirow{2}{*}{$F^{-1}_{\mathrm{crit}}$} & \multirow{2}{*}{$F_{\mathrm{crit}}$}\tabularnewline
\cline{4-7}
 &  &  & $G=0$ & $F$ & $G\neq0$ & $F$ &  & \tabularnewline
\hline 
\hline 
colorimeter & 237 & 0.16 & 0.23 & 0.72 & 0.25 & 0.67 & 0.77 & 1.29\tabularnewline
\hline 
display & 67 & 0.16 & 0.24 & 0.634 & 0.29 & 0.53 & 0.61 & 1.63\tabularnewline
\hline 
surface & 233 & 0.13 & 0.11 & 1.19 & 0.12 & 1.06 & 0.77 & 1.30\tabularnewline
\hline 
BFD-P & 131 & 0.12 & 0.085 & 1.43 & 0.099 & 1.23 & 0.71 & 1.41\tabularnewline
\hline 
\end{tabular}
\end{table}

\begin{table}
\centering{}\caption{F-tests for $d^{2}_{\mathrm{rms}}$between RieLE1 and RieLE2 (dataset
optimal parameters). For the \textquotedblleft surface\textquotedblright{}
sets RieLE2 is significantly better than RieLE1, for the others they
are not significantly different.}\label{tab:F-tests-RieLE}
\begin{tabular}{|c|c|c|c|c|c|c|}
\hline 
group & $N$ & RieLE1 & RieLE2 & $F$ & $F^{-1}_{\mathrm{crit}}$ & $F_{\mathrm{crit}}$\tabularnewline
\hline 
\hline 
colorimeter & 237 & 0.14 & 0.13 & 0.92 & 0.77 & 1.29\tabularnewline
\hline 
display & 67 & 0.090 & 0.074 & 0.82 & 0.61 & 1.63\tabularnewline
\hline 
surface & 233 & 0.16 & 0.12 & 0.74 & 0.77 & 1.30\tabularnewline
\hline 
BFD-P & 131 & 0.16 & 0.11 & 0.70 & 0.71 & 1.41\tabularnewline
\hline 
dark surround & 196 & 0.099 & 0.11 & 1.06 & 0.76 & 1.33\tabularnewline
\hline 
\end{tabular}
\end{table}

\begin{table}
\caption{Group and generic parameter values used in Fig.\ref{fig:results}.
$k_{2+}$ and $k_{4+}$ are the values for positive excursions w.r.t.
the adapting point and $k_{2-}$ and $k_{4-}$ for negative ones.
All numbers have been rounded to two significant digits. $N$ is the
total number of color points in each group. Only the parameters $k_{0-4}$
scale with $d\sigma$. Since the \textquotedblleft BFD-P\textquotedblright{}
dataset is part of the \textquotedblleft surface\textquotedblright{}
sets these 131 color points are included in the 233 \textquotedblleft surface\textquotedblright{}
color points. In total 733 color points are taken into account. In
addition to the scaling factors $F_{g}$ for each group, where we
arbitrarily set $F_{g}=1$ for the \textquotedblleft surface\textquotedblright{}
group, we applied an overall factor $F=7.533$ to match our generic
$d\sigma$ with those of CIEDE2000. Details of the scaling procedure
followed are given in Appendix~\ref{sec:Scaling-of-the}.}\label{tab:parameter-values}

\centering{}%
\begin{tabular}{|c||c|c|c||c|c||c|c|}
\hline 
 & \multicolumn{5}{c||}{\begin{cellvarwidth}[t]
\centering
group parameters\\
$k_{i,\mathrm{group}}$
\end{cellvarwidth}} & \multicolumn{2}{c|}{\begin{cellvarwidth}[t]
\centering
generic param. \\
$k_{i,gen}$
\end{cellvarwidth}}\tabularnewline
\hline 
groups & \begin{cellvarwidth}[t]
\centering
dark \\
sur.
\end{cellvarwidth} & colorim. & disp. & surf. & BFD-P & \begin{cellvarwidth}[t]
\centering
dark sur.,\\
colorim.,\\
disp.
\end{cellvarwidth} & surf.\tabularnewline
\hline 
$N$ & 196 & \multicolumn{1}{c|}{237} & \multicolumn{1}{c||}{67} & \multicolumn{1}{c|}{233} & (131) & 500 & 233\tabularnewline
\hline 
$F_{g}$ & 0.174 & 0.367 & 0.1 & 1 & - & - & -\tabularnewline
\hline 
\hline 
$k_{0}\left[10^{-3}\right]$ & 16 & 8.2 & 29 & 4.9 & 9.1 & 22 & 37\tabularnewline
\hline 
$k_{1}\left[10^{-3}\right]$ & 1.4 & 0.64 & 2.3 & 0.24 & 1.2 & \multicolumn{2}{c|}{1.8}\tabularnewline
\hline 
$k_{2+}\left[10^{-3}\right]$ & 9.8 & 6.4 & 21 & 5.3 & 30 & 15 & 40\tabularnewline
\hline 
$k_{2-}\left[10^{-3}\right]$ & 1.9 & 4.3 & 9.7 & 4 & 22 & 5.8 & 30\tabularnewline
\hline 
$k_{3}\left[10^{-3}\right]$ & 21 & 9.5 & 40 & 4.7 & 25 & 28 & 35\tabularnewline
\hline 
$k_{4+}\left[10^{-3}\right]$ & 27 & 14 & 35 & 8.1 & 47 & 32 & 61\tabularnewline
\hline 
$k_{4-}\left[10^{-3}\right]$ & 0 & 0 & 0 & 0 & 1.8 & \multicolumn{2}{c|}{0}\tabularnewline
\hline 
\hline 
$k_{12}$ & 6.9 & 0 & 8.1 & 53 & 0.8 & 6.6 & 53\tabularnewline
\hline 
$k_{23}$ & 0.14 & 0.21 & 0.30 & 0.56 & 0.56 & 0.22 & 0.56\tabularnewline
\hline 
\end{tabular}
\end{table}

\section{Perceptual coordinates for RieLE2}\label{sec:Perceptual-coordinates-for}

RieLE2 as presented in §~\ref{sec:The-development-of} and using
the coordinates $\left(\ln Y,l,s\right)$ can explicitly be written
as 
\begin{equation}
{d\sigma}^{2}=\frac{1}{f^{2}_{c}}\left[\left(\frac{f_{c}}{f_{A}}\frac{dY}{k_{0}Y}+\delta_{12}\sqrt{\tilde{g}_{22}}dl\right)^{2}+\left(\sqrt{\tilde{g}_{22}}dl+\delta_{23}\sqrt{\tilde{g}_{33}}ds\right)^{2}+\tilde{g}_{33}{ds}^{2}\right]\label{eq:ReLE2explicit}
\end{equation}

where $f_{A}$ and $f_{c}$ are similar functions of $Y$, defined
in Eq.(\ref{eq:fcdef-1}), the $\delta$'s have been defined in Eq.(\ref{eq:delta12})
and Eq,$(\ref{eq:delta23_1})$ and the chromatic tensor elements have
been defined as
\begin{equation}
\sqrt{\tilde{g}_{22}}=\frac{1}{\tilde{\psi}_{T}}=\frac{1}{k_{1}l_{E}+k_{2\pm}\left|l-l_{a}\right|}\label{eq:psiT~}
\end{equation}
\begin{equation}
\sqrt{\tilde{g}_{33}}=\frac{1}{\tilde{\psi}_{D}}=\frac{1}{k_{3}s_{E}+k_{4\pm}\left|s-s_{a}\right|}\label{eq:psiD~}
\end{equation}

We define a base lightness function by
\begin{equation}
d\tilde{L}=\frac{f_{c}}{f_{A}}\frac{dY}{k_{0}Y}=\sqrt{\frac{Y+Y_{c}}{Y+Y_{A}}}\frac{dY}{k_{0}Y}\label{eq:baseL}
\end{equation}

This can be integrated analytically (see Appendix~\ref{sec:The-lightness-functions}).

Since all quantities between brackets in (\ref{eq:ReLE2explicit})
are total differentials we define then conformally-Euclidean coordinates
by
\begin{equation}
dA=d\tilde{L}+\delta_{12}\sqrt{\tilde{g}_{22}}dl\label{eq:Aeq}
\end{equation}
\begin{equation}
dl_{c}=\sqrt{\tilde{g}_{22}}dl+\delta_{23}\sqrt{\tilde{g}_{33}}ds\label{eq:lceq}
\end{equation}
\begin{equation}
ds_{c}=\sqrt{\tilde{g}_{33}}ds\label{eq:sceq}
\end{equation}

These equations can indeed be integrated, starting with Eq.(\ref{eq:sceq})
\begin{align}
s_{c}\left(s-s_{a}\right) & =\int^{s-s_{a}}_{0}\frac{dx}{\tilde{\psi}_{D}}\label{eq:scresult}\\
 & =\int^{s-s_{a}}_{0}\frac{dx}{k_{3}s_{E}+k_{4\pm}\left|x\right|}\nonumber \\
 & =\frac{1}{\pm k_{4\pm}}\ln\left(1+\frac{k_{4\pm}}{k_{3}}\frac{\left|s-s_{a}\right|}{s_{E}}\right)\nonumber 
\end{align}

where the $\pm$ corresponds with $s\gtrless s_{a}$ and $k_{4\pm}$
is different for these regions.

Taking into account that $\delta_{23}=k_{23}$ is a constant we define
an intermediate quantity $\Phi_{T}=l_{c}-\delta_{23}s_{c}$ and rewrite
Eq.(\ref{eq:lceq}) as
\begin{equation}
d\Phi_{T}=dl_{c}-\delta_{23}ds_{c}=\sqrt{\tilde{g}_{22}}dl
\end{equation}

and this equation can then be integrated in the same way
\begin{align}
\Phi_{T}\left(l-l_{a}\right) & =\int^{l-l_{a}}_{0}\frac{dx}{\tilde{\psi}_{T}}\label{eq:PhiTresult}\\
 & =\int^{l-l_{a}}_{0}\frac{dx}{k_{1}l_{E}+k_{2\pm}\left|x\right|}\nonumber \\
 & =\frac{1}{\pm k_{2\pm}}\ln\left(1+\frac{k_{2\pm}}{k_{1}}\frac{\left|l-l_{a}\right|}{l_{E}}\right)\nonumber 
\end{align}

Finally from Eq.(\ref{eq:Aeq}) we define
\begin{equation}
A=\tilde{L}+\varphi\left(l-l_{a}\right)
\end{equation}

where
\begin{align}
\varphi\left(l-l_{a}\right) & =k_{12}\int^{l-l_{a}}_{0}\frac{x^{2}}{\tilde{\psi}_{T}}dx\\
 & =k_{12}\int^{l-l_{a}}_{0}\frac{x^{2}}{k_{1}l_{E}+k_{2\pm}\left|x\right|}dx\nonumber 
\end{align}

is also readily integrated with the result
\begin{equation}
\varphi\left(l-l_{a}\right)=\pm\frac{k_{12}}{k_{2\pm}}\frac{k^{2}_{1}}{k^{2}_{2\pm}}l^{2}_{E}\left[\ln\left(1+\frac{k_{2\pm}}{k_{1}}\frac{\left|l-l_{a}\right|}{l_{E}}\right)-\frac{k_{2\pm}}{k_{1}}\frac{\left|l-l_{a}\right|}{l_{E}}\left(1-\frac{1}{2}\frac{k_{2\pm}}{k_{1}}\frac{\left|l-l_{a}\right|}{l_{E}}\right)\right]
\end{equation}

Using Eq.(\ref{eq:PhiTresult}) this can also be written as a function
of $\Phi_{T}=l_{c}-\delta_{23}s_{c}$
\begin{equation}
\varphi\left(\Phi_{T}\right)=\frac{k_{12}}{k_{2\pm}}\frac{k^{2}_{1}}{k^{2}_{2\pm}}l^{2}_{E}\left[k_{2\pm}\Phi_{T}\pm\frac{1}{2}\left(e^{k_{2\pm}\left|\Phi_{T}\right|}-1\right)\left(e^{k_{2\pm}\left|\Phi_{T}\right|}-3\right)\right]\label{eq:varphidef}
\end{equation}

The small argument limits of these expressions are third order and
given by
\begin{equation}
\varphi\left(\Phi_{T}\rightarrow0\right)=\frac{1}{3}k_{12}k^{2}_{1}l^{2}_{E}\Phi^{3}_{T}\left(1+\frac{3}{4}k_{2\pm}\left|\Phi_{T}\right|+...\right)\label{eq:varphiresultsmall}
\end{equation}
\begin{equation}
\varphi\left(l-l_{a}\rightarrow0\right)=\frac{1}{3}\frac{k_{12}}{k_{1}l_{E}}\left(l-l_{a}\right)^{3}\left(1-\frac{3}{4}\frac{k_{2\pm}}{k_{1}l_{E}}\left|l-l_{a}\right|+...\right)
\end{equation}

We note that $k_{2\pm}\left|\Phi_{T}\right|$ and the complete expression
between brackets in Eq.(\ref{eq:varphidef}) is scale invariant and
$\varphi$ thus scales as $k^{-1}_{2\pm}$ or $k^{-1}_{1}$.

Using again that $\delta_{23}=k_{23}$ is a constant we can write
$\Phi_{T}=\sqrt{1+k^{2}_{23}}l_{cR}$ where $\left(l_{cR},s_{cR}\right)$
are rotated coordinates w.r.t. $\left(l_{c},s_{c}\right)$
\begin{equation}
\begin{bmatrix}l_{cR}\\
s_{cR}
\end{bmatrix}=\frac{1}{\sqrt{1+k^{2}_{23}}}\begin{bmatrix}1 & -k_{23}\\
k_{23} & 1
\end{bmatrix}\begin{bmatrix}l_{c}\\
s_{c}
\end{bmatrix}
\end{equation}

We have thus obtained a conformally Euclidean LE
\begin{equation}
d\sigma^{2}=\frac{1}{f^{2}_{c}}\left(dA^{2}+dl^{2}_{c}+ds^{2}_{c}\right)=\frac{1}{f^{2}_{c}}\left(dA^{2}+dl^{2}_{cR}+ds^{2}_{cR}\right)
\end{equation}
where $f^{2}_{c}=1+Y_{c}/Y$ and $Y$ is a function of $A$ and $\Phi_{T}=l_{c}-k_{23}s_{c}=\sqrt{1+k^{2}_{23}}l_{cR}$.
Considering a subspace with fixed luminance $Y$, this subspace is
thus Euclidean, but it is not a geodesic subspace, meaning that a
geodesic in this subspace is not a geodesic in the enclosing space.

\begin{figure}
\begin{centering}
\includegraphics[width=10cm]{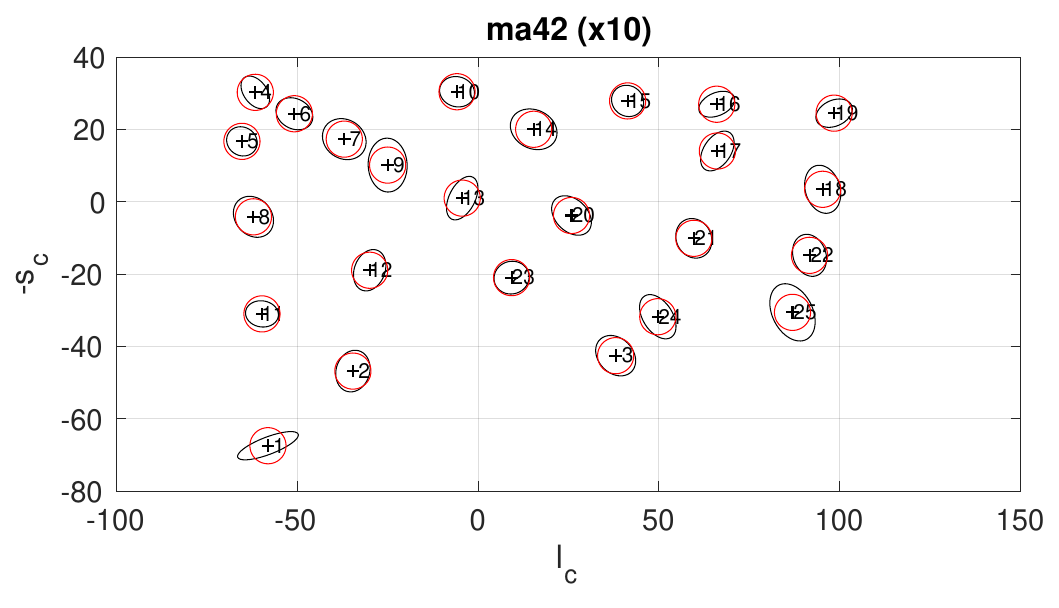}
\par\end{centering}
\caption{Chromatic ellipses (enlarged $\times10$) for the iconic MacAdam
set \textquotedblleft ma42\textquotedblright{} with dataset optimal
parameters ($d_{rms}=0.25$). Experimental ellipses are in black and
the (circular) model ones in red.}\label{fig:ma42 ellipses}
\end{figure}

\begin{figure}
\begin{centering}
\subfloat[]{\includegraphics[width=10cm]{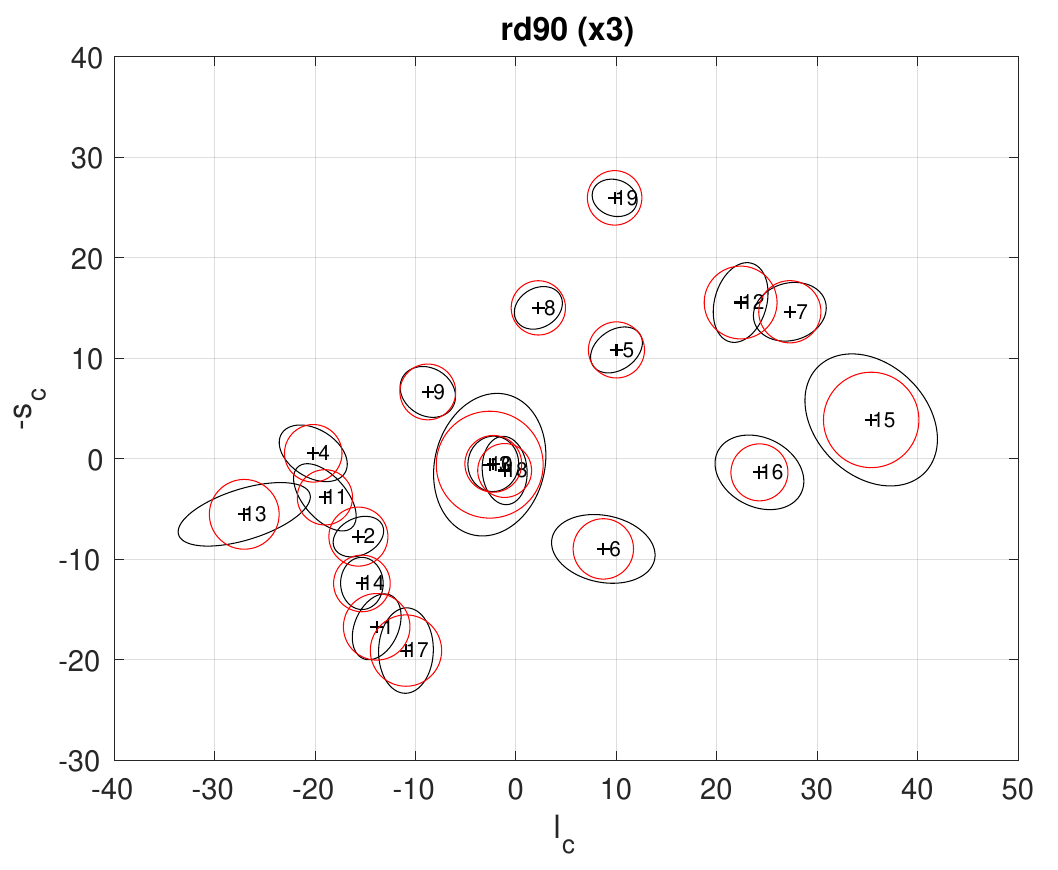}

}
\par\end{centering}
\begin{centering}
\subfloat[]{\includegraphics[totalheight=8cm]{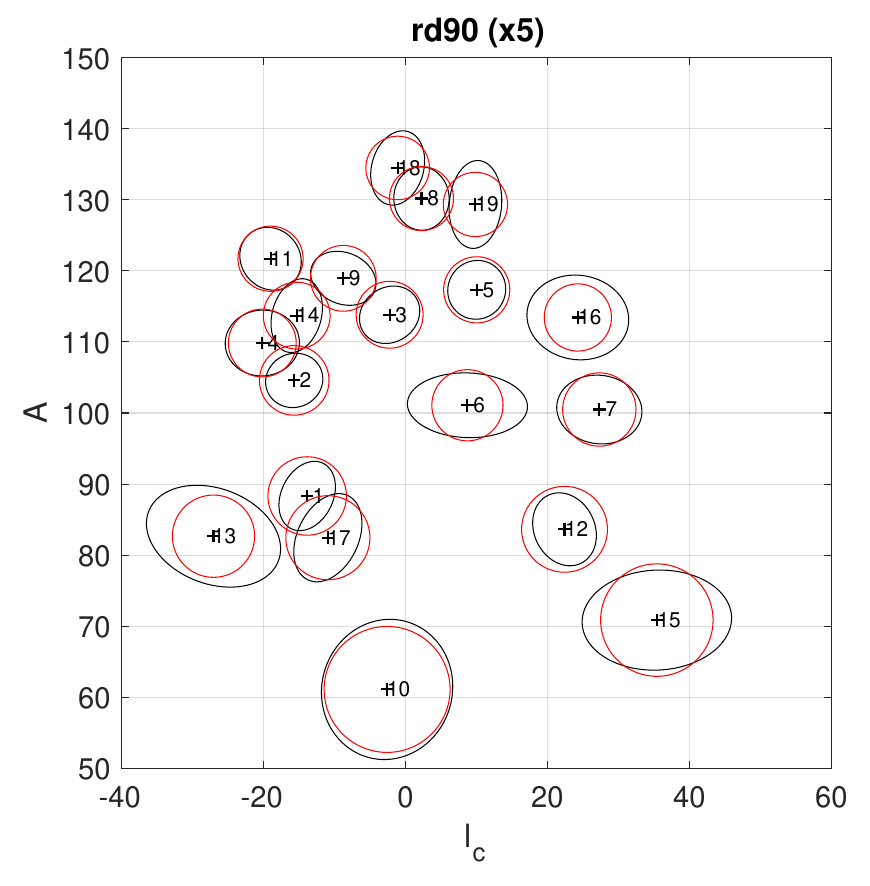}

}\hspace{1cm}\subfloat[]{\includegraphics[totalheight=8cm]{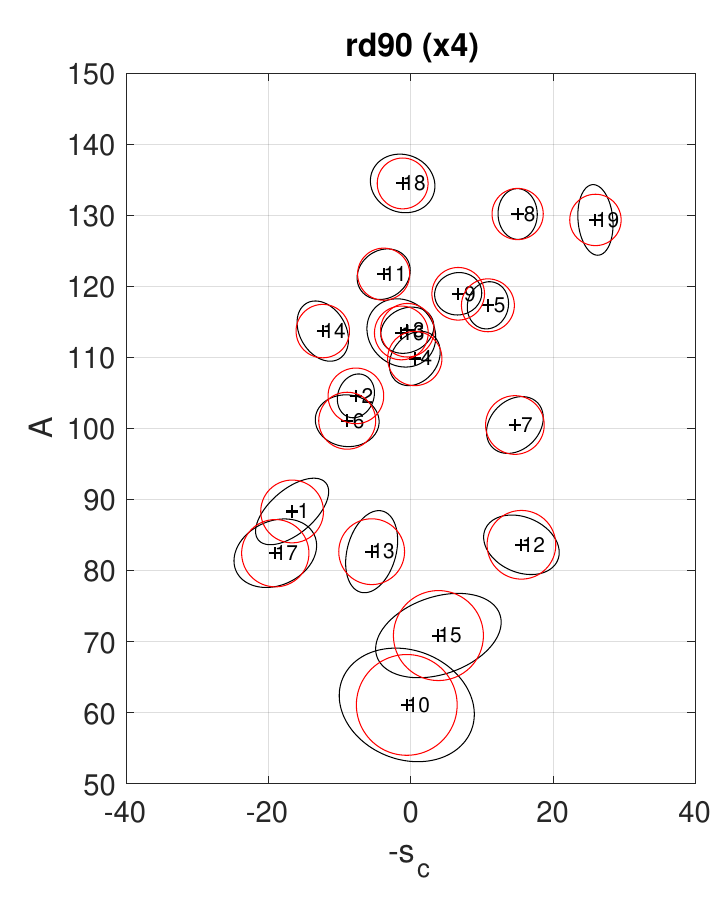}

}
\par\end{centering}
\caption{Cross-sections of the ellipsoids in the 3 main planes for \textquotedblleft surface\textquotedblright{}
dataset rd90 and for the optimal dataset parameters. The ellipses
are enlarged (a) $\times3$, (b) $\times5$ and (c) $\times4$ ($d_{rms}=0.28$).
Experimental ellipses are in black and the (circular) model ones in
red.}\label{fig:rd90 ellipses}
\end{figure}

\begin{figure}
\begin{centering}
\subfloat[]{\includegraphics[totalheight=8cm]{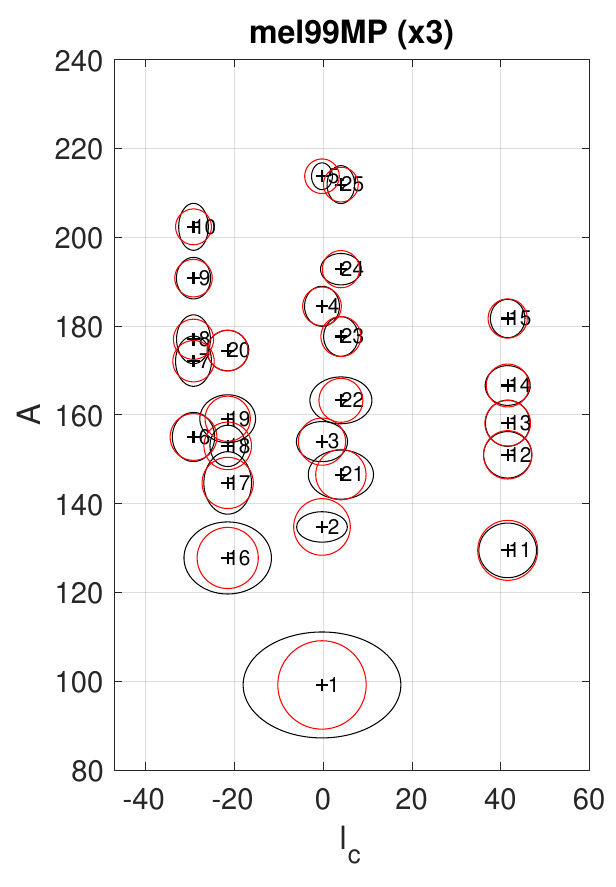}

}\hspace{1cm}\subfloat[]{\includegraphics[totalheight=8cm]{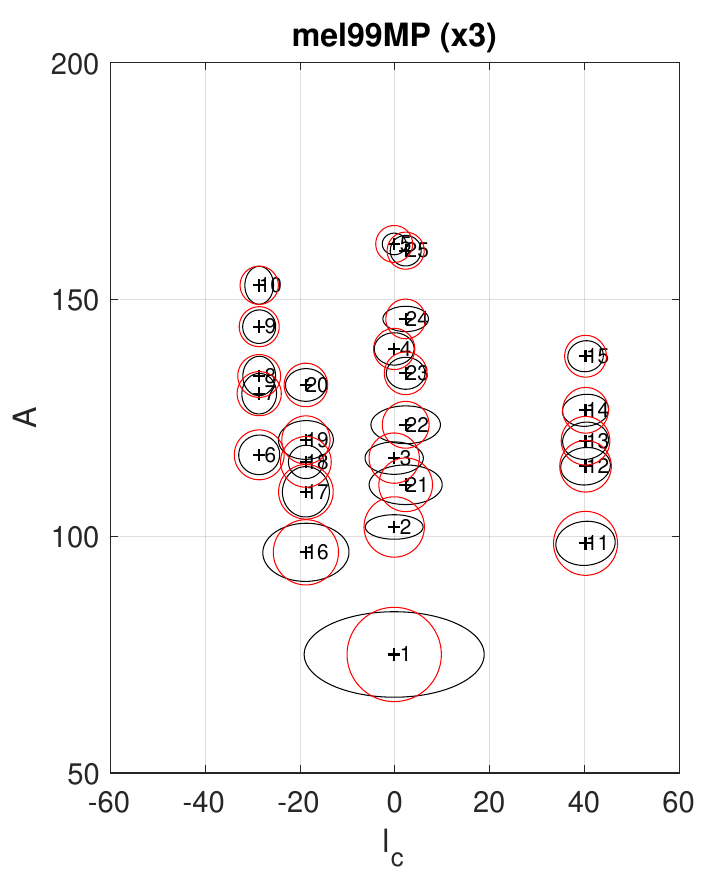}

}
\par\end{centering}
\begin{centering}
\subfloat[]{\includegraphics[totalheight=8cm]{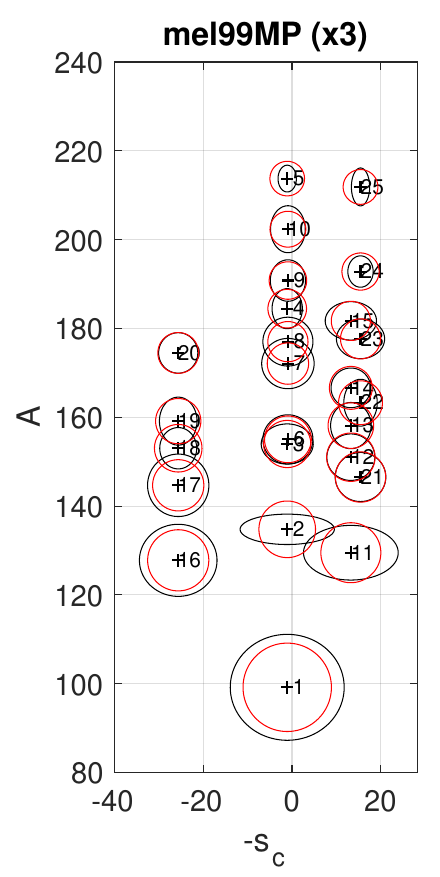}

}\hspace{1cm}\subfloat[]{\includegraphics[totalheight=8cm]{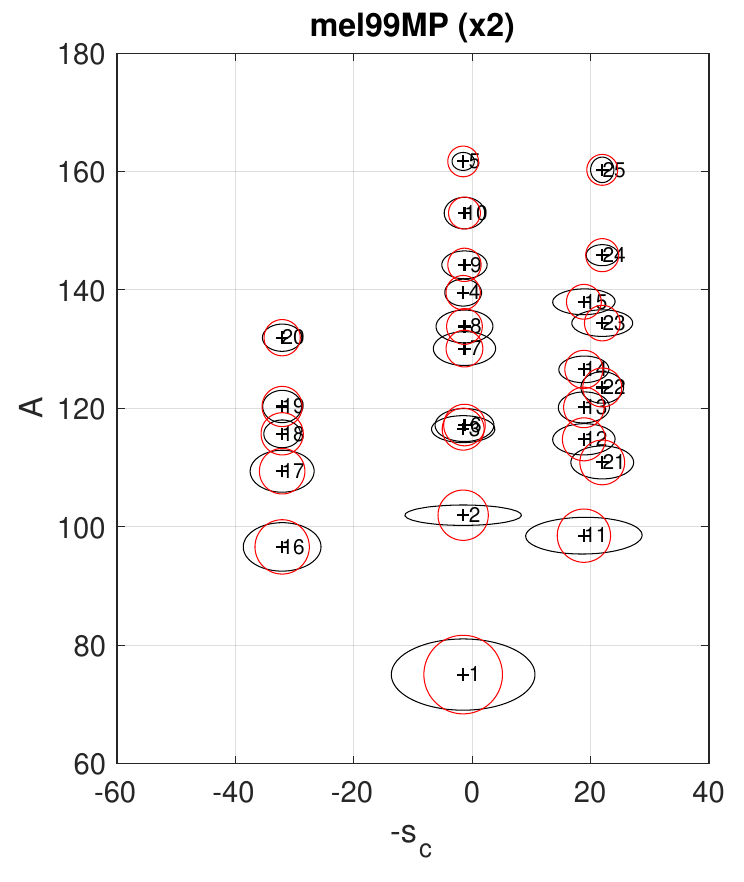}

}
\par\end{centering}
\caption{Cross-sections of the threshold ellipsoids along the planes containing
the achromatic axis for the \textquotedblleft display\textquotedblright{}
dataset mel99MP. The left column (a)(c) shows the results for optimal
dataset parameters ($d_{rms}=0.25$). and the right column (a)(d)
for the \textquotedblleft display\textquotedblright{} generic parameters
($d_{rms}=0.36$). The ellipses are enlarged $\times3$ or $\times2$.
Experimental ellipses are in black and the (circular) model ones in
red.}\label{fig:mel99MP ellipses}
\end{figure}

Examples of ellipses in those coordinates $\left(A,l_{c},s_{c}\right)$
are shown in Fig.\ref{fig:ma42 ellipses}-\ref{fig:mel99MP ellipses}
for respectively the MacAdam set (ma42, optimal dataset parameters),
the rd90 set (optimal dataset parameters) and mel99MP set (optimal
dataset and generic parameters). In these coordinates the model ellipses
are thus all circular with a radius only dependent on the luminance
$Y$.

Obviously the coordinates $\left(A,l_{c},s_{c}\right)$ depend on
the parameters of the model $k_{0-4}$, which are scalable, and the
non-scalable $k_{12}$ and $k_{23}$. When defining those coordinates
we scale the scalable parameters by the corresponding size $d\sigma$
so that always comparable numbers are obtained. The perceptual coordinates
$\left(A,l_{c},s_{c}\right)$ are thus defined, depending on the situation,
with the scalable parameters $k_{i,\mathrm{set}}/d\sigma\approx k_{i,\mathrm{group}}F_{g}F\approx k_{i,\mathrm{gen}}$
($i=0\cdots4$, see Appendix~\ref{sec:Scaling-of-the} for the precise
meaning).

These coordinates $\left(A,l_{c},s_{c}\right)$ are similar to the
CIE $\left(L^{*},a^{*},b^{*}\right)$ coordinates which is illustrated
in Fig.\ref{fig:Relation-between-(a)} where for the dataset BFD-P
(see §~(\ref{subsec:Surface-samples-data})) these coordinates are
compared using the generic parameters $k_{i,\mathrm{gen}}$. The correlation
coefficients are respectively $\rho_{\mathrm{gen}}\left(A,L^{*}\right)=0.9811$,
$\rho_{\mathrm{gen}}\left(l_{c},a^{*}\right)=0.9569$ and $\rho_{\mathrm{gen}}\left(-s_{c},b^{*}\right)=0.9529$.
With the optimal parameters for this dataset $k_{i,\mathrm{set}}$,
the correlation coefficients are respectively $\rho_{\mathrm{set}}\left(A,L^{*}\right)=0.9959$,
$\rho_{\mathrm{set}}\left(l_{c},a^{*}\right)=0.9538$ and $\rho_{\mathrm{se}t}\left(-s_{c},b^{*}\right)=0.9532$.
$\left(A,L^{*}\right)$ differ the most with the generic parameters
but agree the most for the dataset parameters. The others are comparable
in both cases with $\left(l_{c},a^{*}\right)$ differing the most,
and $-s_{c}$ levels off $\approx30$ (since $s$ is strictly constrained
to positive values). There are some differences between the hue angles
around $90^{{^\circ}}$ and especially $270^{{^\circ}}$ although
they correlate very well with $\rho_{\mathrm{gen}}\left(\mathrm{hue}\right)=0.9961$
and $\rho_{\mathrm{set}}\left(\mathrm{hue}\right)=0.9954$, where
we excluded the color points with $C^{*}_{ab}<2$. Since $Y$ and
$f_{c}$ depend on $\Phi_{T}=l_{c}-k_{23}s_{c}$ only, and not on
$l_{c}$ and $s_{c}$ separately, and since $k_{23}$ is a constant,
it would be advantageous to rotate the chromatic plane but then the
correspondence with the CIE $a^{*},b^{*},h^{*}$ would be lost. The
spectrum locus in the unrotated perceptual coordinates is shown in
Supplement~1.
\begin{center}
\begin{figure}
\begin{centering}
\subfloat[]{\includegraphics[totalheight=5cm]{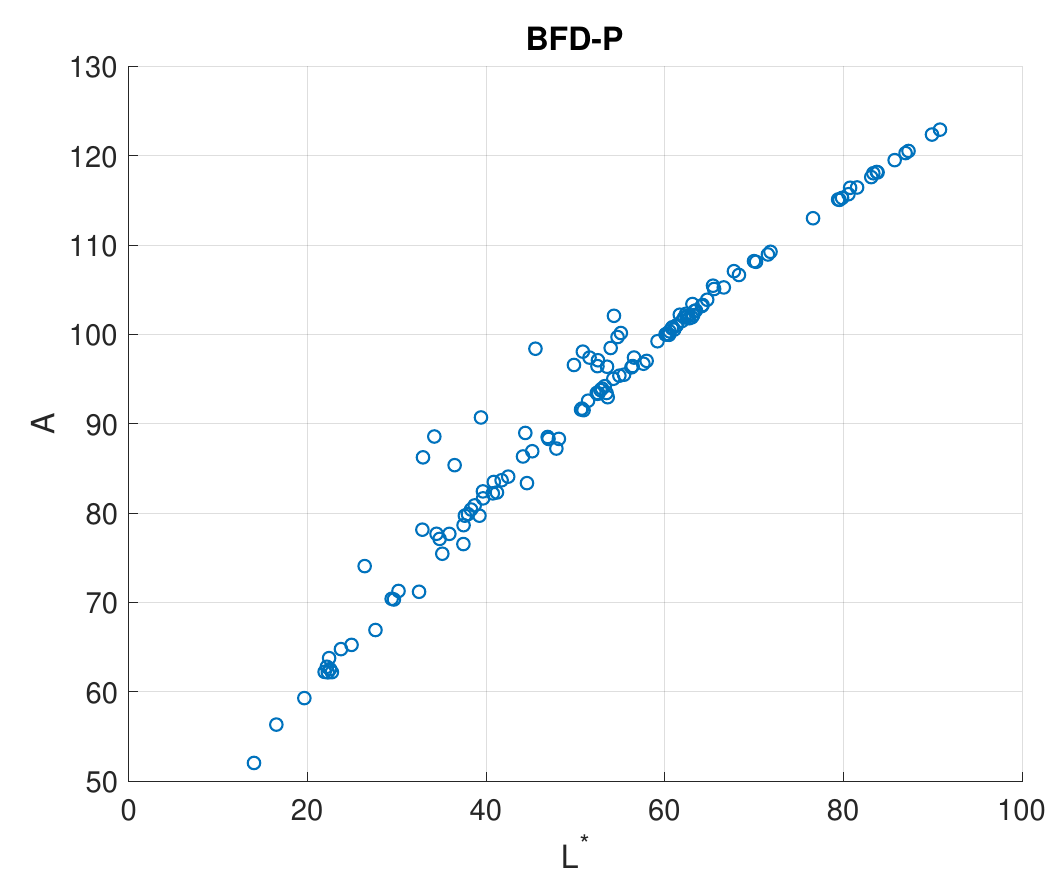}

}\hspace{0.5cm}\subfloat[]{\includegraphics[totalheight=5cm]{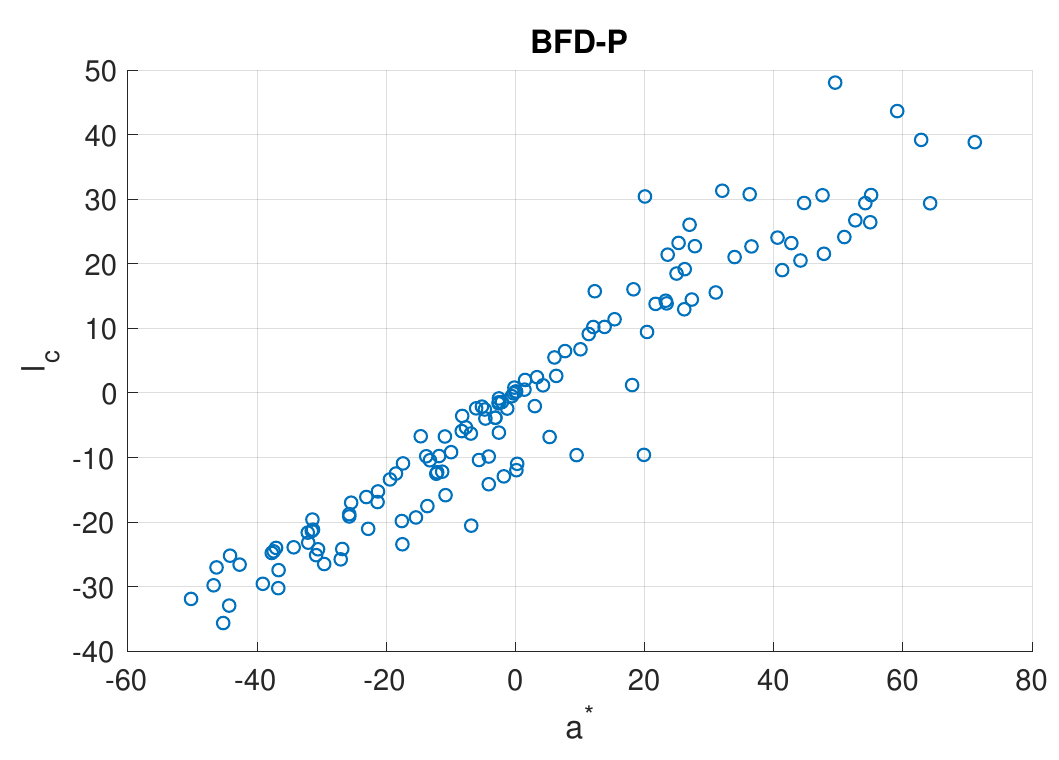}

}
\par\end{centering}
\begin{centering}
\subfloat[]{\includegraphics[totalheight=5cm]{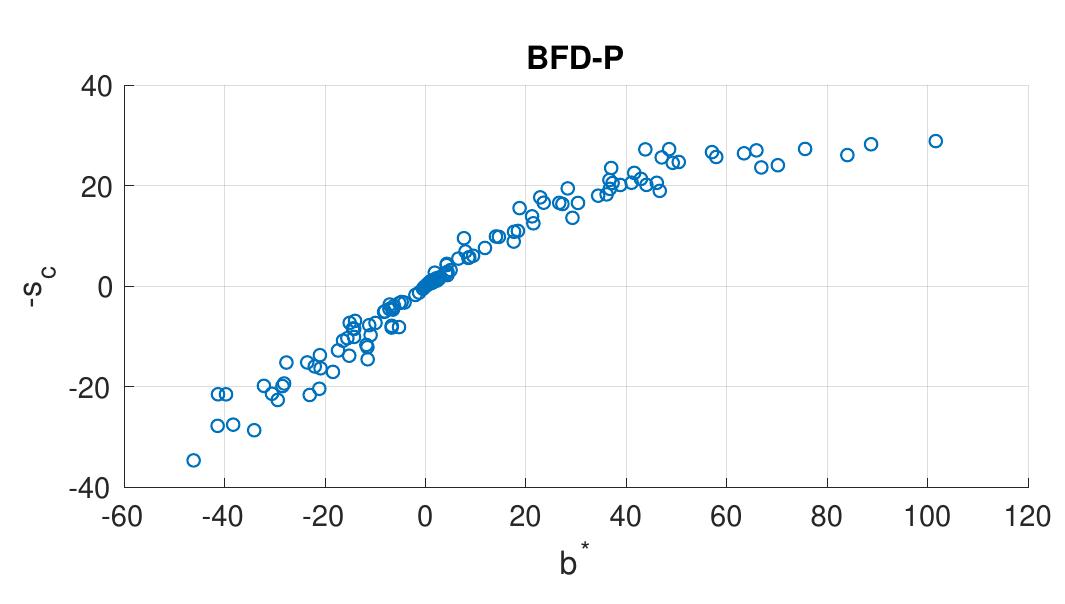}

}
\par\end{centering}
\begin{centering}
\subfloat[]{\begin{centering}
\includegraphics[totalheight=6cm]{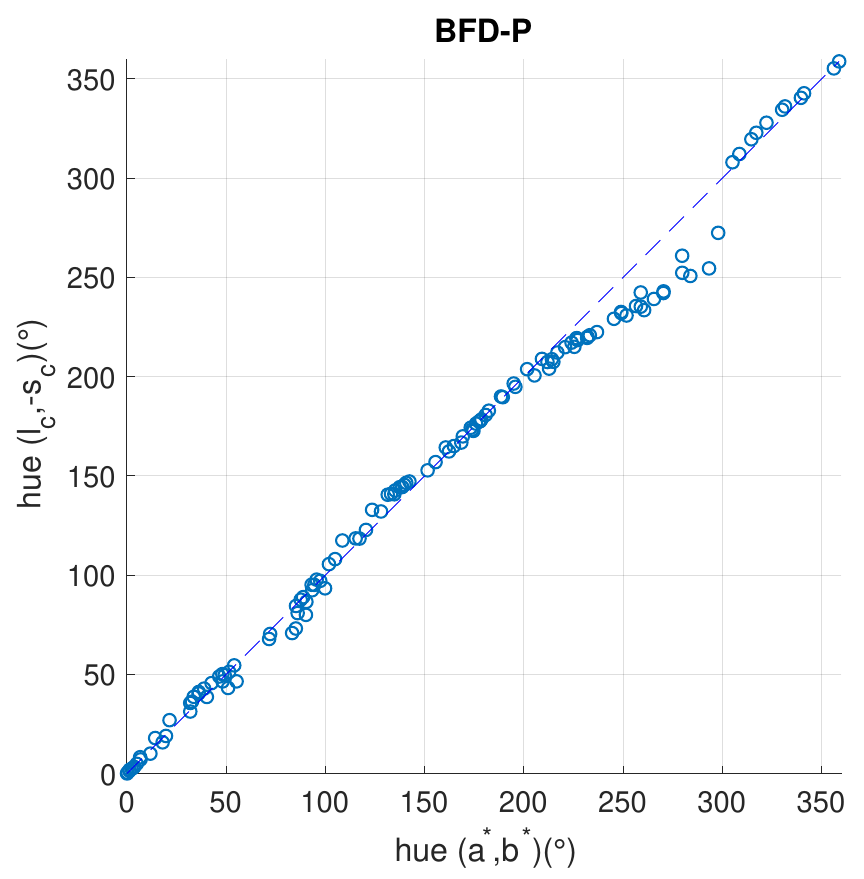}
\par\end{centering}
}
\par\end{centering}
\caption{Relation between (a) $A$ and $L^{*}$, (b) $l_{c}$ and $a^{*}$
(c) $-s_{c}$ and $b^{*}$ and (d) the corresponding hue angles for
the BFD-P \textquotedblleft surface color\textquotedblright{} dataset.
The results have been obtained with the surface generic parameters
$k_{i,\mathrm{gen}}$.}\label{fig:Relation-between-(a)}
\end{figure}
\par\end{center}

\section{Conclusions}

Starting from our previously defined line element which was developed
along lines originally set by Friele, and considering a wider range
of datasets, we simplified it resulting into a conformally-Euclidean
line element, meaning that up to a multiplicative factor the line
element is Euclidean. For surface color datasets this LE performs
significantly better than the original one, comparing optimized datasets
with an F-test. For the other media (colorimeter, display, dark surround)
there is no significant difference. This underlying Euclidean line
element defines natural perceptual coordinates $\left(A,l_{c},s_{c}\right)$
which are very similar to the CIE coordinates $\left(L^{*},a^{*},b^{*}\right)$.
Although the multiplicative factor is a complicated function of these
perceptual coordinates, defining completely the curved geometry of
the color space, it depends on the luminance only and has a very simple
relation. For these coordinates the threshold ellipsoids are all spherical
with a radius only depending on the luminance. We tested the line
element against 38 published datasets comprising a total of 733 color
points evaluated by 231-242 observers. We derived optimal parameters
for each dataset, for each group of datasets (colorimeter, display,
surface color, dark surround) and proposed generic parameters for
the (colorimeter, display, dark surround) collection on the one hand
and the surface colors on the other hand. Comparison between the experimental
ellipsoids and the ones defined by the line element was done with
an error measure based on a comparison of the eigenvalues of the ellipsoids.
For the surface color datasets our LE performs on average equally
well as the CIEDE2000 difference equation (with CIEDE2000 still significantly
better for the surface subset BFD-P supposed one does not implement
the $a^{*}\rightarrow a^{\prime}$ transform), but for the colorimeter
and display datasets it is significantly better and for the dark surround
datasets the CIEDE2000 equation cannot be applied. We have thus developed
a line element using physiologically sound coordinates, that is widely
applicable and with a simple and clear geometry. The sole variable
affecting the curvature of the human color space according to this
line element is the retinal illuminance.\clearpage{}

\bibliographystyle{plainurl}
\bibliography{color,papers2,books}

\appendix

\section{Scaling of the line element}\label{sec:Scaling-of-the}

The datasets (see Appendix~\ref{sec:Overview-of-the}) have been
grouped mainly according to the medium used into four groups: experiments
done with ``colorimeters'', with ``displays'', with ``surface
colors'' and a special group for experiments with a ''dark surround''.
Each dataset comprises experimentally determined tensors $g_{i}$
describing ellipsoids with unknown sizes $d\sigma_{i}$ which for
particular coordinates $x$ (column vector) are given by
\begin{equation}
{d\sigma_{i}}^{2}=dx^{\prime}\cdot g_{i}\cdot dx\label{eq:experimental}
\end{equation}

where the prime denotes the matrix transpose and the $dx$ denotes
the variations around the color point considered. Since our difference
metric is coordinate independent the specific coordinates $x$ used
are not important and being also scale invariant optimally matched
model ellipsoids will have the same average size as the experimental
ones.

The parameters of our (model) line element are twofold: homogeneous
parameters which scale with the average size $d\sigma$ of the ellipsoid
($k_{0-4}$) on the one hand and rotation parameters $k_{12}$ and
$k_{23}$ on the other hand, which do not scale with $d\sigma$. The
latter are of course also chosen optimally but they play no role for
scaling and we can leave them out for the moment. Since the average
size $d\sigma=\left\langle d\sigma_{i}\right\rangle $ of the experimental
ellipsoids is not known we initially find optimal parameters $k_{\mathrm{set}}$
for each dataset with $\left\langle d\sigma_{i}\right\rangle =1$
that is with a model equation, written formally as
\begin{equation}
1=dx^{\prime}\cdot\left[\frac{1}{k^{2}_{i,\mathrm{set}}}\right]\cdot dx\label{eq:model_dataset}
\end{equation}

emphasizing that the tensor elements are proportional with $k^{-2}_{\mathrm{set}}$.
Actually we should write this as
\begin{equation}
{d\sigma}^{2}=dx^{\prime}\cdot\left[\frac{{d\sigma}^{2}}{k^{2}_{i,\mathrm{set}}}\right]\cdot dx\label{eq:model_dataset-scaled}
\end{equation}

We now describe the method we have used to find estimates for those
average dataset sizes $d\sigma$. To this end we apply the same optimization
process, which was applied separately to each dataset, to all color
points of a group but by applying a different scale factor $d\sigma_{\mathrm{group}}$
for the color points of each dataset, yielding a model equation
\begin{equation}
d\sigma^{2}_{\mathrm{group}}=dx^{\prime}\cdot\left[\frac{1}{k^{2}_{i,\mathrm{group}}}\right]\cdot dx\label{eq:model_group}
\end{equation}

where $k_{i,\mathrm{group}}$ are now the same for all datasets in
this group but the $d\sigma_{\mathrm{group}}$ are different for each
dataset. It follows from eq.(\ref{eq:model_dataset}) and eq.(\ref{eq:model_group})
that $\left\langle k_{i,\mathrm{set}}/d\sigma_{\mathrm{group}}\right\rangle \approx k_{i,\mathrm{group}}$.

Although we have now obtained an estimate for the relative sizes $d\sigma_{\mathrm{group}}$
within this particular group we must still take into account an unknown
scale factor $F_{g}$ at the level of the group, meaning we can transform
$d\sigma_{\mathrm{group}}\rightarrow d\sigma_{\mathrm{group}}/F_{g}$
and $k_{\mathrm{group}}\rightarrow k_{\mathrm{group}}F_{g}$ , with
the same matching error. The ``real'' sizes $d\sigma_{\mathrm{group}}/F_{g}$
have been found by comparing the group parameters $k_{\mathrm{group}}$
between the different groups and minimizing their variance. Because
this scaling depends on the particular parameter considered, we have
applied it to the two most important ones available for all datasets
$k_{1}$ and $k_{3}$ and have taken their average.

We also have matched the datasets with the LE of the CIEDE2000 difference
equation but since this LE has fixed parameters the match yields an
average optimal size
\begin{equation}
d\sigma^{2}_{\mathrm{CIE}}=dx^{\prime}\cdot g_{\mathrm{CIE}}\cdot dx\label{eq:model_CIE}
\end{equation}

We have then applied a final overall scale factor $F$ to all datasets
from all groups to align our final sizes $d\sigma$ with those found
with the CIEDE2000 line element and this defines then the generic
parameters $k_{\mathrm{gen}}$ with
\begin{equation}
d\sigma^{2}=dx^{\prime}\cdot\left[\frac{1}{k^{2}_{i,\mathrm{gen}}}\right]\cdot dx\label{eq:model_generic}
\end{equation}

where $d\sigma_{\mathrm{group}}\approx F_{g}Fd\sigma$ and conversely
$F_{g}Fk_{i,\mathrm{group}}\approx k_{i,\mathrm{gen}}$.

The values of $d\sigma_{\mathrm{group}}$ and of these final $d\sigma$
have been listed in the tables in Appendix~\ref{sec:Overview-of-the},
the values of $k_{i,\mathrm{group}}$ and $k_{i,\mathrm{gen}}$ and
the scale factors $F_{g}$ and $F$ are shown in Table.\ref{tab:parameter-values}
and an overview of all parameters is shown in Fig.\ref{fig:kplots}.

Summarizing the different parameters are related by $k_{i,\mathrm{set}}/d\sigma\approx k_{i,\mathrm{group}}F_{g}F\approx k_{i,\mathrm{gen}}$
and for these parameters we get indeed from eqs.~(\ref{eq:model_dataset})(\ref{eq:model_group})(\ref{eq:model_generic})
\begin{align}
d\sigma^{2} & =dx^{\prime}\cdot\left[\frac{d\sigma^{2}}{k^{2}_{i,\mathrm{set}}}\right]\cdot dx\nonumber \\
d\sigma^{2} & =dx^{\prime}\cdot\left[\frac{1}{F^{2}_{g}F^{2}k^{2}_{i,\mathrm{group}}}\right]\cdot dx\label{eq:model_all}\\
d\sigma^{2} & =dx^{\prime}\cdot\left[\frac{1}{k^{2}_{i,\mathrm{gen}}}\right]\cdot dx\nonumber 
\end{align}

\section{The lightness functions}\label{sec:The-lightness-functions}

The base lightness defined in (\ref{eq:baseL}) is much simpler if
$Y_{A}=Y_{c}$ with then
\begin{equation}
k_{0}\tilde{L}\left(Y_{A}=Y_{c}\right)=\ln Y\label{eq:k0Lspecial}
\end{equation}

where we arbitrarily set the zero for $Y=1$. More generally the integral
can still be found analytically but is more cumbersome yielding after
some manipulations
\begin{multline}
k_{0}\tilde{L}=a\ln Y+\left(1-a\right)\ln\left(1+\frac{Y}{Y_{A}}\right)-c\\
+2\left[\ln\left(1+\sqrt{\frac{Y+Y_{c}}{Y+Y_{A}}}\right)-a\ln\left(\sqrt{\frac{Y+Y_{c}}{Y+Y_{A}}}+a\right)\right]\label{eq:koLgeneral}
\end{multline}

where $a^{2}=Y_{c}/Y_{A}$ and where the constant $c=2\left[\ln\left(1+a\right)-a\ln\left(2a\right)\right]$.
This expression reduces to (\ref{eq:k0Lspecial}) for $Y_{A}=Y_{c}$
($a=1$), but if $Y_{c}\neq Y_{A}$ the zero point is slightly below
1. A fairly good approximation is obtained by replacing the 2nd line
in (\ref{eq:koLgeneral}) by its limit $Y\rightarrow\infty$ yielding
\begin{equation}
k_{0}\tilde{L}^{\infty}=a\ln Y+\left(1-a\right)\ln\left(1+\frac{Y}{Y_{A}}\right)+c^{\prime}\label{eq:k0Lapprox}
\end{equation}

where the constant is given by
\begin{equation}
c^{\prime}=2\left[a\ln a-\left(1+a\right)\ln\frac{1+a}{2}\right]
\end{equation}

Whereas Eq.(\ref{eq:k0Lspecial}) is easily inverted, Eqs.(\ref{eq:koLgeneral})
and (\ref{eq:k0Lapprox}) can only be inverted numerically. However
Eq.(\ref{eq:k0Lapprox}) has 2 simple asymptotes meeting at $Y=Y_{A}$
and given by
\begin{equation}
k_{0}\tilde{L}^{\infty}\approx\begin{cases}
a\ln Y+c^{\prime} & Y<Y_{A}\\
\ln Y+c^{\prime\prime} & Y>Y_{A}
\end{cases}\label{eq:k0Lasymptotes}
\end{equation}

where $c^{\prime\prime}=c^{\prime}-(1-a)\ln Y_{A}$. With these expressions
good initial guesses can be obtained for inverting Eq.(\ref{eq:koLgeneral}).

The function $\tilde{L}$ defined above is not equivalent with the
CIEDE2000 lightness function since we must still multiply with the
conformal factor and thus
\[
dL=\frac{1}{f_{c}}d\tilde{L}
\]

With (\ref{eq:baseL}) we find
\[
k_{0}dL=\frac{dY}{\sqrt{Y\left(Y+Y_{A}\right)}}
\]

which is easily integrated and yields
\begin{equation}
k_{0}L\left(Y\right)=\ln\left[1+2\left(\frac{Y}{Y_{A}}+\sqrt{\frac{Y}{Y_{A}}\left(\frac{Y}{Y_{A}}+1\right)}\right)\right]\label{eq:k0Lfull}
\end{equation}

which becomes zero for $Y=0$, but to avoid the logarithmic divergence
can be limited to $Y>1$.

\section{The conversion from CIE $\left(Y,x,y\right)$ coordinates to conformally-Euclidean
coordinates}

The conversion from CIE $\left(X,Y,Z\right)$ coordinates to the fundamental
cone responses $\left(L,M,S\right)$ is done by a matrix transformation
which can generally be written as 
\begin{equation}
\begin{bmatrix}L\\
M\\
S
\end{bmatrix}=\begin{bmatrix}\alpha & \beta & -\gamma\\
-\alpha & 1-\beta & \gamma\\
0 & 0 & \delta
\end{bmatrix}\begin{bmatrix}X\\
Y\\
Z
\end{bmatrix}\label{eq:XYZtoLMS}
\end{equation}

where $\alpha,\beta,\gamma$ are constants defined by $\alpha^{-1}=x_{p}/y_{p}-x_{d}/y_{d}$
, $\beta=-\alpha x_{d}/y_{d}$ and $\gamma=\alpha x_{t}/z_{t}$ and
where $(x_{p,}y_{p})$, $(x_{d},y_{d})$ and $(x_{t},y_{t})$ are
the coordinates of the protanopic, deuteranopic and tritanopic ``confusing''
points in the CIE chromaticity space, with the conditions $x_{p}+y_{p}=1$,
$x_{d}+y_{d}=1$ and $y_{t}=0$. For the Stockman-Sharpe fundamentals
$\alpha=0.1453$, $\beta=0.5899$ and $\gamma=0.0274$. Finally, following
Wyszecki and Stiles \cite{Wyszecki:1982oh} (page 615) $\delta=0.0192$
is defined so that for monochromatic light with wavelength $\lambda_{0}=418.1\,\mathrm{nm}$
$s\left(\lambda_{0}\right)=1$ where the MacLeod-Boynton chromaticities
are defined by $l=L/Y$ and $s=S/Y$. Combining these results we get
\begin{equation}
\begin{bmatrix}l\\
s
\end{bmatrix}=\begin{bmatrix}\alpha+\gamma & \beta+\gamma\\
-\delta & -\delta
\end{bmatrix}\begin{bmatrix}x/y\\
1
\end{bmatrix}+\begin{bmatrix}-\gamma\\
\delta
\end{bmatrix}\frac{1}{y}\label{eq:xytols}
\end{equation}

The compressed MacLeod-Boynton chromaticities are given by (\ref{eq:scresult})
and (\ref{eq:PhiTresult})
\begin{equation}
s_{c}=\frac{1}{\pm k_{4\pm}}\ln\left(1+\frac{k_{4\pm}}{k_{3}}\frac{\left|s-s_{a}\right|}{s_{E}}\right)
\end{equation}
\begin{equation}
l_{c}=\frac{1}{\pm k_{2\pm}}\ln\left(1+\frac{k_{2\pm}}{k_{1}}\frac{\left|l-l_{a}\right|}{l_{E}}\right)+\delta_{23}s_{c}
\end{equation}

and the achromatic coordinate is given by
\begin{equation}
A=\tilde{L}\left(Y\right)+\varphi\left(l_{c}-\delta_{23}s_{c}\right)
\end{equation}

where $\tilde{L}$ is defined in (\ref{eq:koLgeneral}) and
\begin{equation}
\varphi\left(\Phi_{T}\right)=\frac{k_{12}}{k_{2\pm}}\frac{k^{2}_{1}}{k^{2}_{2\pm}}l^{2}_{E}\left[k_{2\pm}\Phi_{T}\pm\frac{1}{2}\left(e^{k_{2\pm}\left|\Phi_{T}\right|}-1\right)\left(e^{k_{2\pm}\left|\Phi_{T}\right|}-3\right)\right]
\end{equation}

Depending on the level considered (dataset, group, generic) we use
parameters $k_{i,\mathrm{set}}$, $k_{i,\mathrm{group}}d\sigma_{\mathrm{group}}$
or $k_{i,\mathrm{gen}}d\sigma$ to define those coordinates. In the
resulting space $\left(A,l_{c},s_{c}\right)$ the threshold ellipsoid
is spherical with a radius depending on the luminance only
\begin{equation}
f_{c}=\sqrt{1+\frac{Y_{c}}{Y}}
\end{equation}

Its derivatives w.r.t. the perceptual coordinates can be calculated
with
\begin{equation}
\frac{\partial f_{c}}{\partial A}=-\frac{k_{0}/2}{1+Y/Y_{c}}\sqrt{1+\frac{Y_{A}}{Y}}
\end{equation}
\begin{equation}
\frac{\partial f_{c}}{\partial l_{c}}=-\varphi^{\prime}\left(\Phi_{T}\right)\frac{\partial f_{c}}{\partial A}
\end{equation}
\begin{equation}
\frac{\partial f_{c}}{\partial s_{c}}=k_{23}\varphi^{\prime}\left(\Phi_{T}\right)\frac{\partial f_{c}}{\partial A}
\end{equation}

where the ``prime'' denotes differentiation w.r.t. the argument.
Threshold ellipsoids defined in $\left(\ln Y,x,y\right)$ can be converted
to $\left(A,l_{c},s_{c}\right)$ by the transformation of the differentials
\[
\begin{bmatrix}dx\\
dy
\end{bmatrix}=\frac{y}{\alpha}\begin{bmatrix}1-x & \frac{\gamma}{\delta}-\frac{\alpha+\gamma}{\delta}x\\
-y & -\frac{\alpha+\gamma}{\delta}y
\end{bmatrix}\begin{bmatrix}dl\\
ds
\end{bmatrix}
\]
\[
\begin{bmatrix}\frac{dY}{Y}\\
dl\\
ds
\end{bmatrix}=\begin{bmatrix}k_{0}\frac{f_{A}}{f_{c}} & -k_{0}\frac{f_{A}}{f_{c}}\delta_{12} & k_{0}\frac{f_{A}}{f_{c}}\delta_{12}\delta_{23}\\
0 & \tilde{\psi}_{T} & -\delta_{23}\tilde{\psi}_{T}\\
0 & 0 & \tilde{\psi}_{D}
\end{bmatrix}\begin{bmatrix}dA\\
dl_{c}\\
ds_{c}
\end{bmatrix}
\]

\section{Overview of the datasets used to test LE's}\label{sec:Overview-of-the}

We only mention the most essential parameters, besides the identifying
information, the number of observers involved or their abbreviations,
the number of color points considered ($N$) and the range of $Y$-values.
More details can be found in the original publications. For the surface
samples the particular medium is also mentioned. Many datasets are
restricted to chromatic thresholds and these are labeled as ``chromatic
only'' (co). The $d\sigma$ columns mention the optimal sizes of
the ellipsoids for the generic parameters listed in Table~\ref{tab:parameter-values},
which has been scaled to match the CIEDE2000 values optimally (coefficient
of determination $R^{2}=0.90$)
\[
d\sigma=\frac{d\sigma_{\mathrm{group}}}{F_{g}F}
\]

where $d\sigma_{\mathrm{group}}$ is the optimal size of a dataset
within a group for the group parameters. Details of the scaling model
used are given in Appendix~\ref{sec:Scaling-of-the}.

\subsection{Colorimeter data}
\begin{center}
\begin{tabular}{|c|c|c|c|c|c|c|c|}
\hline 
\multirow{1}{*}{code} & \multirow{1}{*}{$N$} & \multirow{1}{*}{co} & \multirow{1}{*}{observer(s)} & \multirow{1}{*}{$Y${[}td{]}} & $d\sigma_{\mathrm{group}}$ & \multicolumn{1}{c|}{$d\sigma$} & \multirow{1}{*}{ref.}\tabularnewline
\hline 
\hline 
ma42 & 25 & $\checkmark$ & 1 & 236.4 & 0.96 & 0.37 & \cite{MacAdam:1942kk}\tabularnewline
\hline 
b57 & 22 &  & 12 & 62.4-133.2 & 0.71 & 0.27 & \cite{Brown:1957ph}\tabularnewline
\hline 
\multirow{3}{*}{wf71} & \multirow{3}{*}{28} &  & GF & \multirow{3}{*}{137.5} & 0.52 & 0.19 & \multirow{3}{*}{\cite{Wyszecki:1971pi}}\tabularnewline
\cline{3-4}\cline{6-7}
 &  &  & AR &  & 0.53 & 0.19 & \tabularnewline
\cline{3-4}\cline{6-7}
 &  &  & GW &  & 0.55 & 0.20 & \tabularnewline
\hline 
\multirow{3}{*}{wf71cdm} & 35 &  & GF & 130.4-151.4 & 1.01 & 0.37 & \multirow{3}{*}{\cite{Wyszecki:1971if}}\tabularnewline
\cline{2-7}
 & 34 &  & AR & 130.4-148 & 0.99 & 0.36 & \tabularnewline
\cline{2-7}
 & 37 &  & GW & 128.9-141.3 & 0.95 & 0.35 & \tabularnewline
\hline 
\end{tabular}
\par\end{center}

\subsection{Display data}
\begin{center}
\begin{minipage}[t]{1\columnwidth}%
\begin{center}
\begin{tabular}{|c|c|c|c|c|c|c|c|}
\hline 
\multirow{1}{*}{code} & \multirow{1}{*}{$N$} & \multirow{1}{*}{co} & \multirow{1}{*}{observer(s)} & \multirow{1}{*}{$Y${[}td{]}} & $d\sigma_{\mathrm{group}}$ & \multicolumn{1}{c|}{$d\sigma$} & \multirow{1}{*}{ref.}\tabularnewline
\hline 
\hline 
mel99 & 25 &  & MP & 25.9-593.7 & 0.84 & 1.12 & \cite{Melgosa:1999pf}\tabularnewline
\hline 
perez00o & 5 & $\checkmark$ & AM & 102.1-574.8 & 0.90 & 1.22 & \cite{Perez:2000bl}\tabularnewline
\hline 
indow92bsM & 5 &  & MI & 149.1-833.8 & 0.43 & 0.58 & \cite{Indow:1992wj}\tabularnewline
\hline 
XuYa2005th & 5 & $\checkmark$ & 8 & 140.5-1106.4 & 0.49 & 0.67 & \cite{Xu:2005vb}\tabularnewline
\hline 
WangXu14gray & 5 & $\checkmark$ & 5 & 143.8-1117.8 & 1.08 & 1.46 & \cite{Wang:2014ot}\tabularnewline
\hline 
ZLuomain & 11 &  & \multirow{2}{*}{26} & 43.6-1571.2 & 1.20 & 1.60 & \multirow{2}{*}{\cite{Luo:2023zp}\cite{Luo:2026hb}}\tabularnewline
\cline{1-3}\cline{5-7}
ZLuotext10 & 11 &  &  & 43.6-1571.2 & 1.26 & 1.68 & \tabularnewline
\hline 
\end{tabular}
\par\end{center}%
\end{minipage}
\par\end{center}

\subsection{Surface samples data}\label{subsec:Surface-samples-data}
\begin{center}
\begin{tabular}{|c|c|c|c|c|c|c|c|c|}
\hline 
\multirow{1}{*}{code} & \multirow{1}{*}{medium} & \multirow{1}{*}{$N$} & \multirow{1}{*}{co} & \multirow{1}{*}{observer(s)} & \multirow{1}{*}{$Y${[}td{]}} & $d\sigma_{\mathrm{group}}$ & \multicolumn{1}{c|}{$d\sigma$} & \multirow{1}{*}{ref.}\tabularnewline
\hline 
\hline 
alderD65 & \begin{cellvarwidth}[t]
\centering
gloss paint,\\
wool
\end{cellvarwidth} & 41 & $\checkmark$ & 19-24 & 26.5-1193.3 & 6.46 & 0.86 & \cite{Alder:2008io}\tabularnewline
\hline 
BFD-P & see §~\ref{subsec:BFD-P-subsets} & 126 & $\checkmark$ &  & 43.8-1976.9 & 5.71 & 0.76 & \cite{Luo:1986fp}\tabularnewline
\hline 
witt90 & gloss paint & 5 &  & 22-24 & 242.2-1903.9 & 1.42 & 0.19 & \cite{Witt:1990ci}\tabularnewline
\hline 
witt99 & paint & 5 &  & 10-14 & 245.3-1944.9 & 1.41 & 0.19 & \cite{Witt:1999ww}\tabularnewline
\hline 
rd90 & gloss paint & 19 &  & 50 & 60.8-2192.8 & 6.43 & 0.86 & \cite{Berns:2009jv}\tabularnewline
\hline 
cheung86 & wool & 5 &  & 20 & 87.2-686.4 & 5.62 & 0.75 & \cite{Cheung:1986rj}\tabularnewline
\hline 
Huang2012b & inkjet print & 17 & $\checkmark$ & 16 & 229.4-1687.9 & 3.79 & 0.51 & \cite{Huang:2012kk}\tabularnewline
\hline 
guan1999esGGH & \multirow{2}{*}{wool} & \multirow{2}{*}{5} & \multirow{2}{*}{$\checkmark$} & \multirow{2}{*}{21} & \multirow{2}{*}{85-561.6} & 3.83 & 0.51 & \multirow{2}{*}{\cite{Guan:1999es}}\tabularnewline
\cline{1-1}\cline{7-8}
guan1999esGGL &  &  &  &  &  & 3.78 & 0.51 & \tabularnewline
\hline 
\end{tabular}
\par\end{center}

\subsection{BFD-P subsets}\label{subsec:BFD-P-subsets}

The BFD-P set is a surface color set comprising a number of many different
sets obtained independently. They were combined by Luo \cite{Luo:1986fp}
in a uniform way and although the data is already contained in §~\ref{subsec:Surface-samples-data}
we thought it useful to handle these subsets also separately, but
some similar data have been grouped.
\begin{center}
\begin{tabular}{|c|c|c|c|c|c|c|c|c|}
\hline 
\multirow{1}{*}{code} & \multirow{1}{*}{medium} & \multirow{1}{*}{$N$} & \multirow{1}{*}{co} & \multirow{1}{*}{observer(s)} & \multirow{1}{*}{$Y${[}td{]}} & $d\sigma_{\mathrm{group}}$ & \multicolumn{1}{c|}{$d\sigma$} & \multirow{1}{*}{reference}\tabularnewline
\hline 
\hline 
CIE & textile & 7 & $\checkmark$ & 20 & 226.4-1613.7 & 1.14 & 0.82 & \multirow{8}{*}{\cite{Luo:1986fp}}\tabularnewline
\cline{1-8}
CISCC & \begin{cellvarwidth}[t]
\centering
matte paint,

textile
\end{cellvarwidth} & 6 & $\checkmark$ & 26-37 & 403.5-1099.3 & 1.04 & 0.74 & \tabularnewline
\cline{1-8}
MCD & \begin{cellvarwidth}[t]
\centering
polyester

thread
\end{cellvarwidth} & 17 & $\checkmark$ & 8 & 250-1469.1 & 1.07 & 0.76 & \tabularnewline
\cline{1-8}
K..W & textile & 23 & $\checkmark$ & 10-30 & 92.7-1927.5 & 1.20 & 0.87 & \tabularnewline
\cline{1-8}
BFD & \begin{cellvarwidth}[t]
\centering
gloss paint,

textile
\end{cellvarwidth} & 41 & $\checkmark$ & 20 & 43.8-1976.9 & 0.99 & 0.71 & \tabularnewline
\cline{1-8}
DF & wool & 12 & $\checkmark$ & 8 & 88.4-1031.4 & 1.06 & 0.76 & \tabularnewline
\cline{1-8}
MMB & ink & 17 & $\checkmark$ & 20 & 152-1288 & 1.05 & 0.76 & \tabularnewline
\cline{1-8}
VVVR & gloss paint & 8 & $\checkmark$ & 14-25 & 55.7-1770.5 & 1.01 & 0.73 & \tabularnewline
\hline 
\end{tabular}
\par\end{center}

\subsection{Dark Surround datasets}

Especially for data obtained with colorimeters, sometimes a uniform
background was replaced by a dark surround, turning the experiment
into one with unrelated colors. Also some such experiments have been
done with displays. We group all these experiments and we assume that
$l_{a}=l_{E}$ and $s_{a}=s_{E}$, that is the adapting color is ``equal
energy white''.
\begin{center}
\begin{minipage}[t]{1\columnwidth}%
\begin{center}
\begin{tabular}{|c|c|c|c|c|c|c|c|c|}
\hline 
\multirow{1}{*}{code} & \multirow{1}{*}{medium} & \multirow{1}{*}{$N$} & \multirow{1}{*}{co} & \multirow{1}{*}{observer(s)} & \multirow{1}{*}{$Y${[}td{]}} & $d\sigma_{\mathrm{group}}$ & \multicolumn{1}{c|}{$d\sigma$} & \multirow{1}{*}{reference}\tabularnewline
\hline 
\hline 
\multirow{2}{*}{bma49} & \multirow{8}{*}{\begin{cellvarwidth}[t]
\centering
colori-

meter
\end{cellvarwidth}} & 38 &  & WRJB & 25- 736 & 0.35 & 0.26 & \multirow{2}{*}{\cite{Brown:1949fp}}\tabularnewline
\cline{3-8}
 &  & 37 &  & DLM & 25-736 & 0.35 & 0.26 & \tabularnewline
\cline{1-1}\cline{3-9}
\multirow{3}{*}{mel90} &  & 12 & $\checkmark$ & AH & 88.6 & 2.24 & 1.66 & \multirow{3}{*}{\cite{Melgosa:1990sf}}\tabularnewline
\cline{3-8}
 &  & 12 & $\checkmark$ & MM & 88.6 & 2.44 & 1.81 & \tabularnewline
\cline{3-8}
 &  & 12 & $\checkmark$ & ME & 88.6 & 2.64 & 1.96 & \tabularnewline
\cline{1-1}\cline{3-9}
\multirow{3}{*}{Ro93} &  & 20 & $\checkmark$ & EG & 257.9 & 3.98 & 2.95 & \multirow{3}{*}{\cite{Romero:1993sz}}\tabularnewline
\cline{3-8}
 &  & 20 & $\checkmark$ & JA & 257.9 & 3.04 & 2.25 & \tabularnewline
\cline{3-8}
 &  & 20 & $\checkmark$ & PL & 257.9 & 3.18 & 2.36 & \tabularnewline
\hline 
\multirow{3}{*}{perez00a} & \multirow{6}{*}{display} & 5 & $\checkmark$ & AM & 189.8-881.6 & 2.68 & 2.00 & \multirow{3}{*}{\cite{Perez:2000bl}}\tabularnewline
\cline{3-8}
 &  & 5 & $\checkmark$ & MP & 187.8- 875.1 & 2.30 & 1.72 & \tabularnewline
\cline{3-8}
 &  & 5 & $\checkmark$ & MM & 189.8- 855.5 & 3.53 & 2.64 & \tabularnewline
\cline{1-1}\cline{3-9}
\multirow{2}{*}{indow92dsM} &  & 5 &  & TI & 225.4-1265.8 & 0.89 & 0.66 & \multirow{3}{*}{\cite{Indow:1992fi}}\tabularnewline
\cline{3-8}
 &  & 5 &  & MI & 225.7-1282.6 & 1.01 & 0.75 & \tabularnewline
\cline{1-1}\cline{3-8}
indow92dsC\footnote{This dataset has an abnormal large optimal error and has not been
taken into account.} &  & 5 &  & MI & 225.7-1282.6 &  &  & \tabularnewline
\hline 
\end{tabular}
\par\end{center}%
\end{minipage}
\par\end{center}
\end{document}


\title{A conformally-Euclidean Line Element for evaluating color differences}

\maketitle
This document provides supplementary information to the manuscript
’A conformally-Euclidean Line Element for evaluating color differences’.

\section{Spectrum locus in RieLE2 perceptual coordinates}\label{sec:Spectrum-locus-in}

\begin{figure}
\begin{centering}
\subfloat[]{\includegraphics[width=10cm]{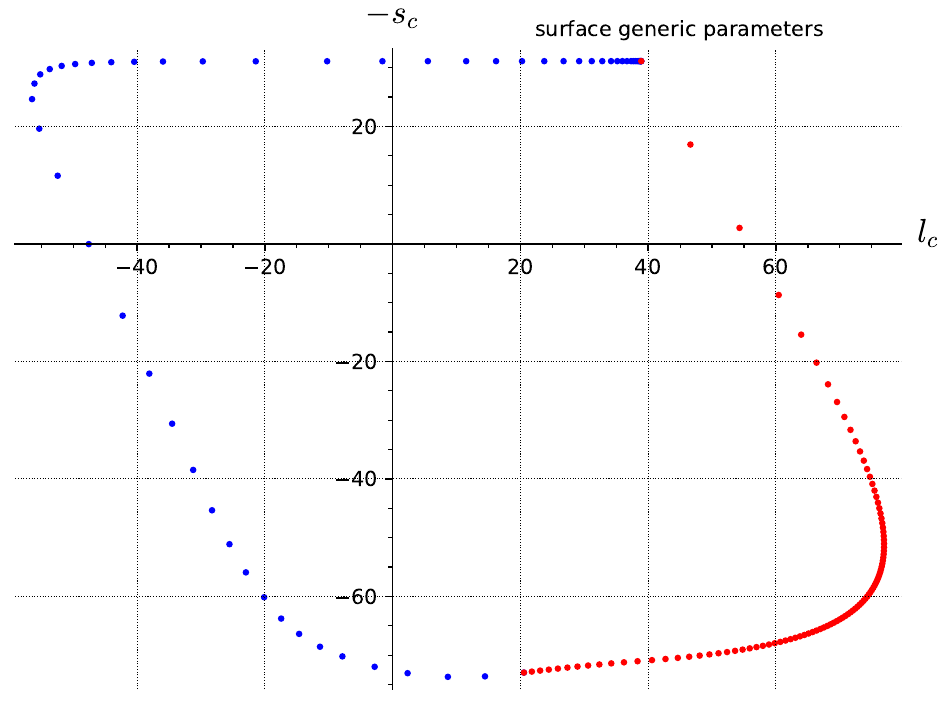}

}
\par\end{centering}
\begin{centering}
\subfloat[]{\includegraphics[width=10cm]{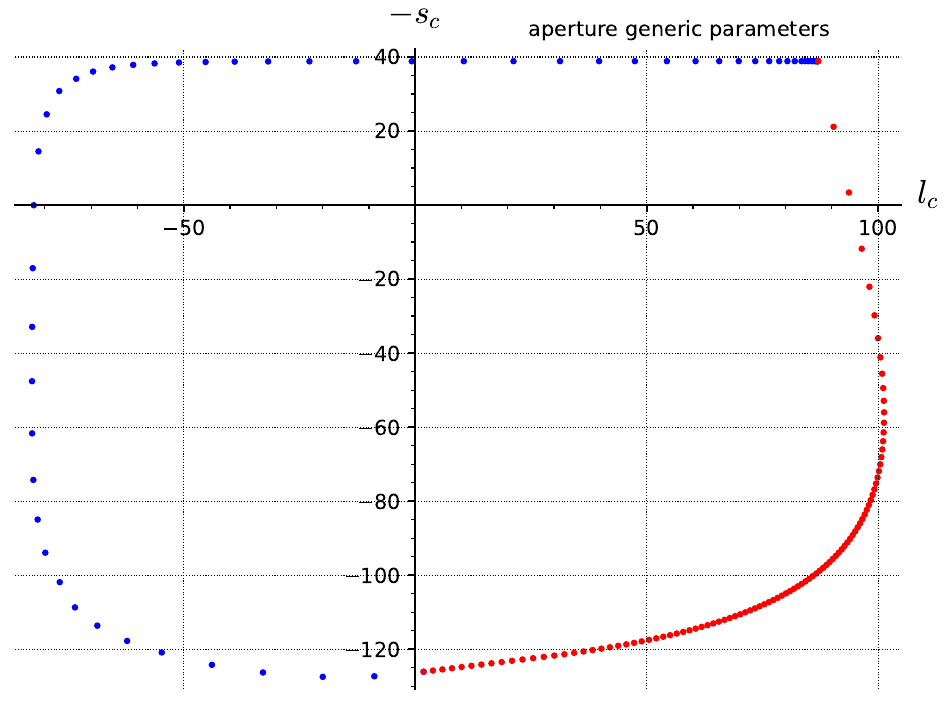}

}
\par\end{centering}
\caption{Spectrum locus and purple line in perceptual coordinates for (a)
\textquotedblleft surface\textquotedblright{} and (b) \textquotedblleft aperture\textquotedblright{}
generic parameters, using the Stockman-Sharp cone fundamentals and
the adapting D65.}\label{fig:Spectrum-locus-and}

\end{figure}

Since the perceptual coordinates depend on the parameters of the line
element they are different for generic ``surface'' parameters and
for ``aperture'' parameters (colorimeter, display and dark surround).
Using the Stockman-Sharpe cone fundamentals \cite{Stockman:2000kq}
the spectrum locus and purple line are shown in Fig.\ref{fig:Spectrum-locus-and}
for an observer adapted to D65.

\section{Comparison between RieLE2 and CIEDE2000 chromatic ellipses}\label{sec:Comparison-between-RieLE2}

\begin{figure}
\begin{centering}
\subfloat[]{\includegraphics[totalheight=8cm]{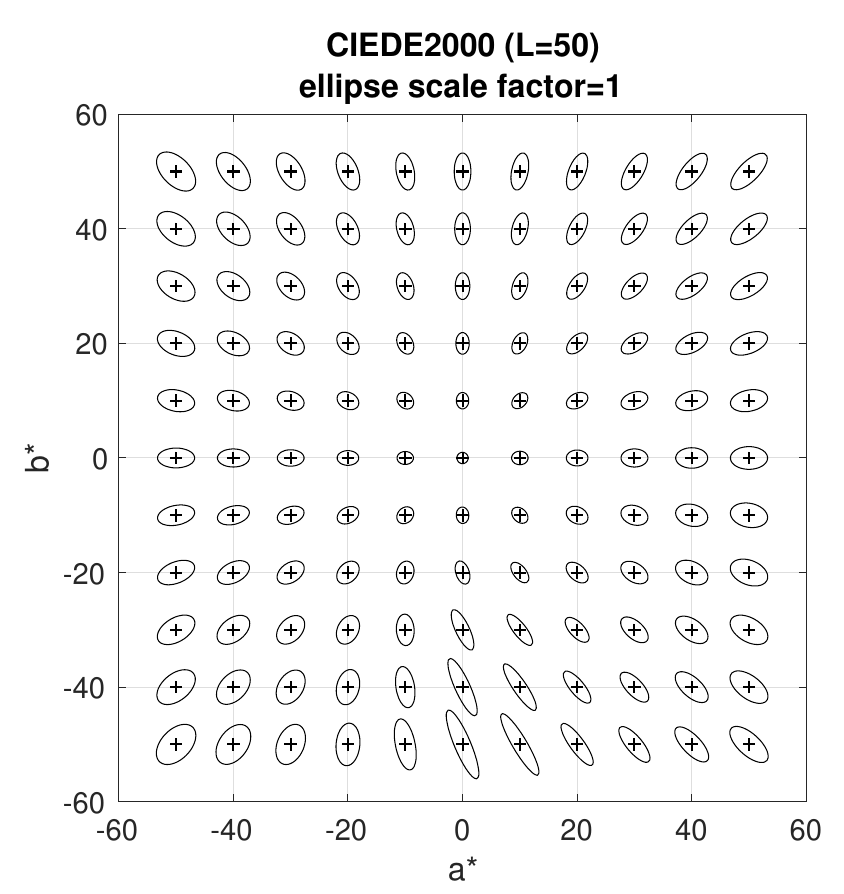}

}$\:$\subfloat[]{\includegraphics[totalheight=8cm]{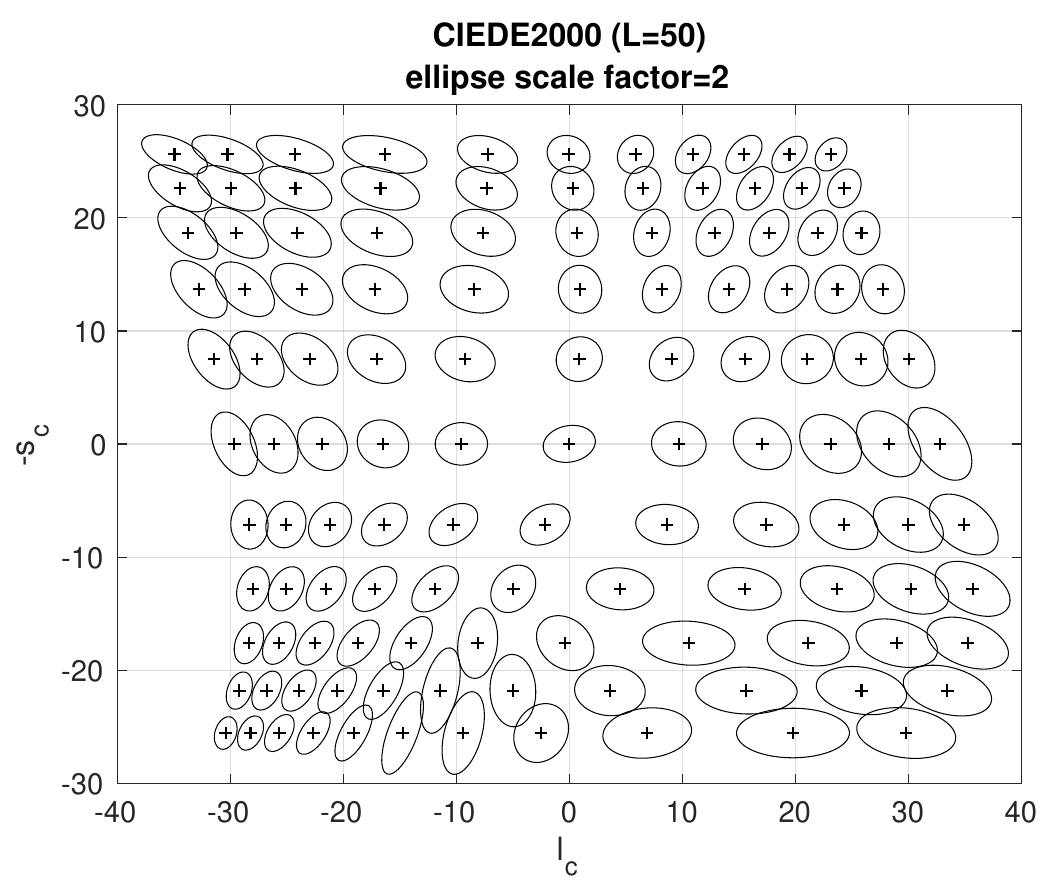}

}
\par\end{centering}
\begin{centering}
\subfloat[]{\includegraphics[totalheight=8cm]{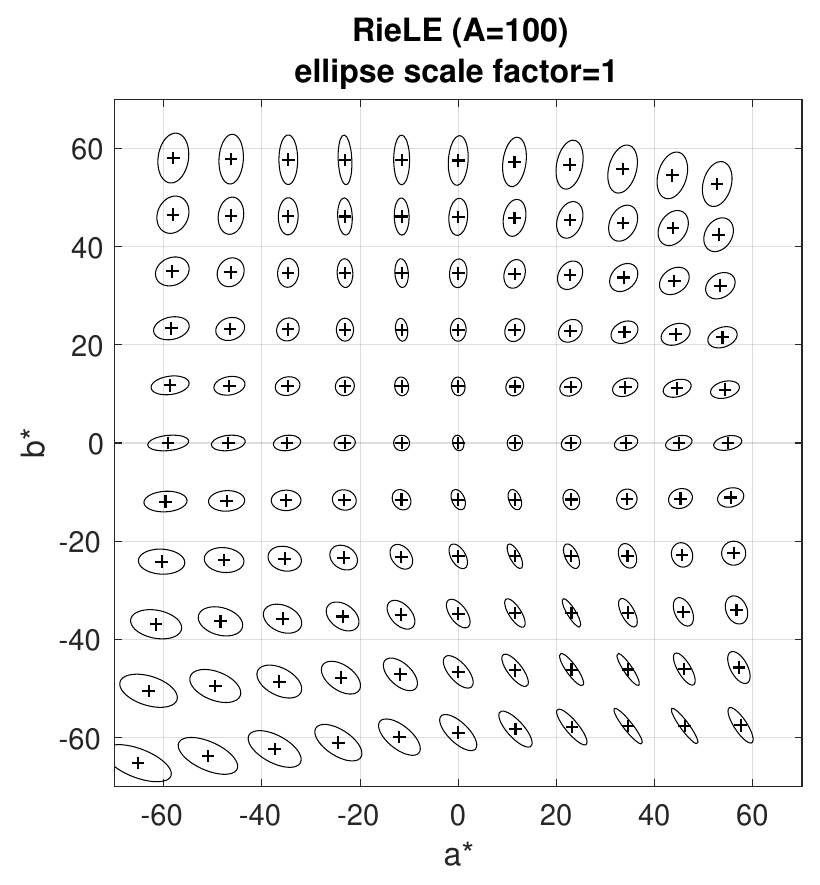}

}$\:$\subfloat[]{\includegraphics[totalheight=8cm]{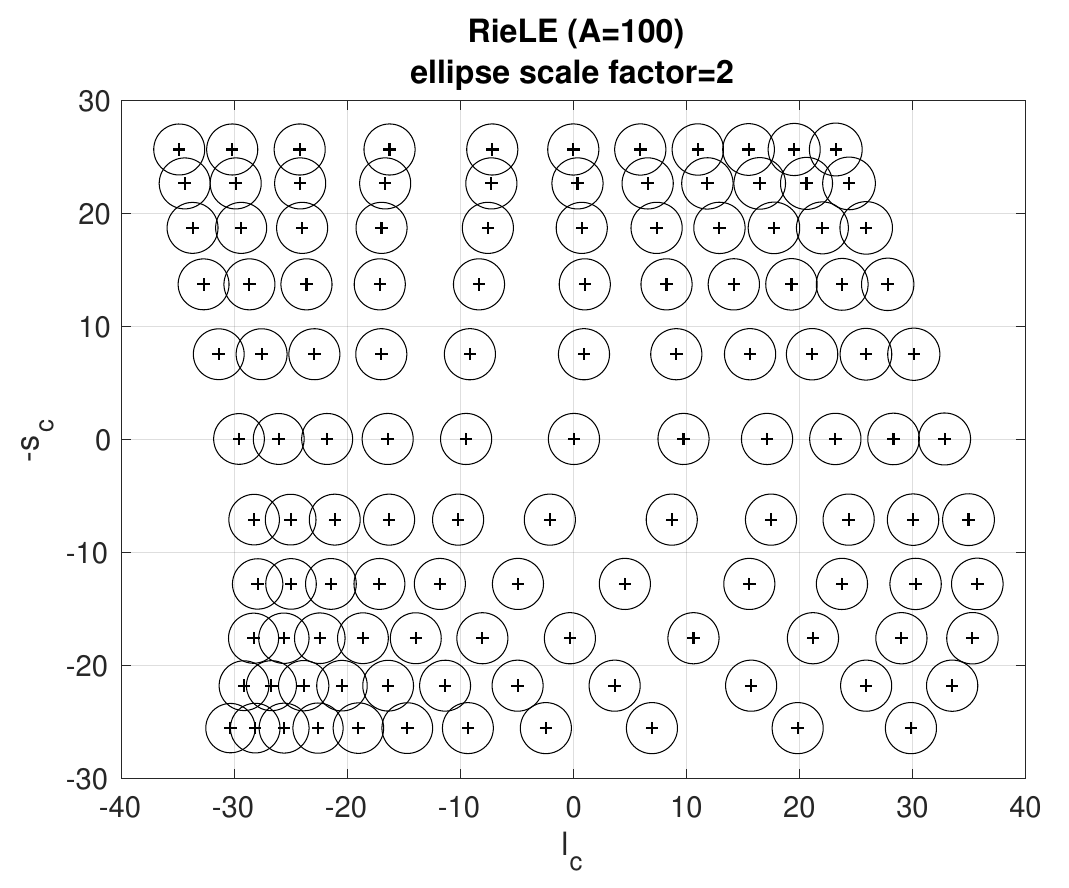}

}
\par\end{centering}
\caption{Comparison between CIEDE2000 ellipses and RieLE2 circles in $\left(L^{*},a^{*},b^{*}\right)$
and $\left(A,l_{c},-s_{c}\right)$.}\label{fig:Comparison-between-CIEDE2000}

\end{figure}

To compare the ellipses predicted by CIEDE2000 and RieLE2 we consider
a regular grid in $\left(a^{*},b^{*}\right)$-space (for $L^{*}=50$)
with the CIEDE2000 ellipses (Fig.\ref{fig:Comparison-between-CIEDE2000}~(a))
and transform them to the $\left(A,l_{c},-s_{c}\right)$-space (Fig.\ref{fig:Comparison-between-CIEDE2000}~(b)).
We then replace these by the RieLE2 circles for $A=100$ (Fig.\ref{fig:Comparison-between-CIEDE2000}~(d))
and transform these circles back to the $\left(L^{*},a^{*},b^{*}\right)$-space
(Fig.\ref{fig:Comparison-between-CIEDE2000}~(c)). As in the main
paper we did not take into account the $a^{*}\rightarrow a^{\prime}$
transformation which is part of CIEDE2000.

\bibliographystyle{unsrturl}
\bibliography{RieLEan-supplement.bib}